%% file: mnras_template.tex
\documentclass[fleqn,usenatbib]{mnras}

\usepackage{newtxtext,newtxmath}

\usepackage[T1]{fontenc}

\DeclareRobustCommand{\VAN}[3]{#2}
\let\VANthebibliography\thebibliography
\def\thebibliography{\DeclareRobustCommand{\VAN}[3]{##3}\VANthebibliography}

\usepackage{graphicx}	
\usepackage{amsmath}	
\usepackage{amssymb}	

\newcommand{\eg}{e.g.,~}
\newcommand{\ie}{i.e.,~}

\title[Observational EBL from {\em HST}--CANDELS]{An observational determination of the evolving extragalactic background light from the multiwavelength {\em HST}/CANDELS survey in the {\em Fermi} and CTA era}

\author[Saldana-Lopez et al.]{
Alberto Saldana-Lopez,$^{1,2}$ \thanks{E-mail: alberto.saldanalopez@unige.ch}
Alberto Dom\'inguez,$^{2}$ \thanks{E-mail: alberto.d@ucm.es}
Pablo G. P\'erez-Gonz\'alez,$^{3}$
\thanks{E-mail: pgperez@cab.inta-csic.es}
\newauthor
Justin Finke,$^{4}$
Marco Ajello,$^{5}$
Joel~R. Primack,$^{6}$
Vaidehi~S. Paliya,$^{7}$
Abhishek Desai,$^{8}$
\\
$^{1}$Observatoire de Gen\`eve, D\'epartement d'Astronomie, Universit\'e de Gen\`eve, 51 Chemin Pegasi, 1290 Versoix, Switzerland\\
$^{2}$ IPARCOS and Department of EMFTEL, Universidad Complutense de Madrid, E-28040 Madrid, Spain\\
$^{3}$ Centro de Astrobiología (CAB, CSIC-INTA), Carretera de Ajalvir km4, E-28850 Torrejón de Ardoz, Madrid, Spain\\
$^{4}$ U.S. Naval Research Laboratory, Code 7653, 4555 Overlook Avenue SW, Washington, DC 20375-5352, USA\\
$^{5}$ Department of Physics and Astronomy, Clemson University, Kinard Lab of Physics, Clemson, SC 29634-0978, USA\\
$^{6}$ Department of Physics, University of California, Santa Cruz, CA 95064, USA\\
$^{7}$ Aryabhatta Research Institute of Observational Sciences (ARIES), Manora Peak, Nainital 263001, India\\
$^{8}$ Dept. of Physics and Wisconsin IceCube Particle Astrophysics Center, University of Wisconsin-Madison, Madison, WI 53706, USA
}

\date{Accepted 2021 August 6. Received 2021 July 23; in original form 2020 December 8}

\pubyear{2021}

\begin{document}
\label{firstpage}
\pagerange{\pageref{firstpage}--\pageref{lastpage}}
\maketitle

\begin{abstract}
The diffuse extragalactic background light (EBL) is formed by ultraviolet (UV), optical, and infrared (IR) photons mainly produced by star formation processes over the history of the Universe, and contains essential information about galaxy evolution and cosmology. Here, we present a new determination of the evolving EBL spectral energy distribution using a novel approach purely based on galaxy data aiming to reduce current uncertainties on the higher redshifts and IR intensities. Our calculations use multiwavelength observations from the UV to the far-IR of a sample of approximately 150~000 galaxies detected up to $z\sim 6$ in the five fields of the Cosmic Assembly Near-Infrared Deep Extragalactic Legacy Survey from the {\it Hubble Space Telescope}. This is one of the most comprehensive and deepest multi-wavelength galaxy datasets ever obtained. These unprecedented resources allow us to derive the overall EBL evolution up to $z\sim 6$ and its uncertainties. Our results agree with cosmic observables estimated from galaxy surveys and $\gamma$-ray attenuation such as monochromatic luminosity densities, including those in the far-IR, and star formation rate densities, also at the highest redshits. Optical depths from our EBL approximation, which will be robust at high redshifts and for $\gamma$-rays up to tens of TeV, will be reported in a companion paper.



\end{abstract}



\newpage

\defcitealias{ChEl01}{CE01}
\defcitealias{DaHe02}{DH02}
\defcitealias{Rieke09}{R09}
\defcitealias{Barro19}{B19}
\defcitealias{Wuyts11}{W11}

\input{1_intro}

\input{2_data}

\input{3_methodology}
\input{4_results}
\input{5_summary}

\section*{Acknowledgements}
ASL acknowledges support from grants FNS-OB5347 and ESP2017-87291-P (AEI/FEDER-UE). AD acknowledges the support of the Ram{\'o}n y Cajal program from the Spanish MINECO. PGP-G acknowledges support from Spanish Government grant PGC2018-093499-BI00. This work has made use of the Rainbow database, which is operated by the Centro de Astrobiología (CAB/INTA), partnered with the University of California Observatories at Santa Cruz (UCO/Lick,UCSC). This work is based in part on observations made with the {\it Spitzer Infrared Space Telescope}, which is operated by the Jet Propulsion Laboratory, California Institute of Technology under a contract with NASA.




\section*{Data availability}
The data products underlying this article are available in the following website: \url{https://www.ucm.es/blazars/ebl}.

\bibliographystyle{mnras}
\bibliography{mnras_template} 






\bsp	
\label{lastpage}
\end{document}

%% file: 1_intro.tex
\section{Introduction}\label{sec:intro} 
Galaxy evolution processes involve the release of radiation to the intergalactic space mainly through two ways: (1) stellar emission and (2) thermal radiation from warmed dust. The former mechanism liberates photons from ultraviolet (UV) to optical and near-infrared (NIR) wavelengths, whereas the latter does both in the mid- and far-infrared range (MIR and FIR, respectively). The major contributors to the UV (sometimes optical) emission are populations of massive and/or young stars; as less massive, older and cooler ones preferentially emit in the NIR \citep[\eg][]{Kippenhahn}. An important fraction of these UV/optical photons produce the excitation of vibrational and rotational levels of ions and molecules in the interstellar and intergalactic medium. Dust grains then heat up, increase their temperature, and emit radiation at lower energies, roughly as a modified black body whose spectrum peaks at wavelengths between 20 and 200~$\mu$m, also in emission bands produced by the relaxation of the excited levels of polycyclic aromatic hydrocarbons \citep[PAHs, \eg][]{DraineISM}. There is also some contribution from active galactic nuclei activity \citep[\eg][]{Hauser01}.


All these photons, which are redshifted to longer wavelengths due to the cosmological expansion, travel through the intergalactic medium (IGM), accumulate in interstellar/intergalactic space, and produce the diffuse background radiation known as the extragalactic background light \citep[EBL, \eg][]{Peebles1993,Peebles1967,Hauser98,primack11,Dwek13}. This background constitutes the second most energetic diffuse radiation field after the cosmic microwave background \citep[\eg][]{Cooray16}. Fundamentally, the evolving spectral energy distribution (SED) of the EBL contains key information about galaxy formation and evolution, such as the cosmic star formation history \citep[CSFH, \eg][]{MD14}, the cosmological expansion of the universe \citep[\eg][]{D13b,D19}, and the cosmic opacity to $\gamma$-rays \citep[\eg][]{Gould66,fazio70,stecker92,D13a}.

Furthermore, a precise and accurate knowledge of the evolving EBL is of fundamental importance for the correct interpretation of the $\gamma$-ray spectra of extragalactic sources \citep[\eg][]{vandenberg19}, the measurement of intergalactic magnetic fields using $\gamma$ rays \citep[\eg][]{neronov10,tavecchio10,essey11,dermer11,arlen14,finke15,ackermann18_highlat}, and the search for exotic physics such as the existence of axion-like particles \citep[\eg][]{sanchezconde09,dominguez11,buehler20} and Lorentz invariance violation \citep[\eg][]{kifune99,abdalla18}. This is also interesting for constraining the redshift of sources at unknown distances \citep[\eg][]{prandini10,abdalla20}.

Measuring the EBL is difficult and may require the convergence of several approaches. Direct measurements, using absolute photometry, are subject to contamination by zodiacal light \citep[\eg][]{Hauser01, Mattila06, Matsuoka11}. This problem makes such estimates challenging, although some attempts have been made at the NIR-to-MIR regime, using instruments onboard the {\it Cosmic Background Explorer} \citep[COBE,][]{Hauser98, Cambresy01}, the {\it Infrared Telescope in Space} \citep[IRTS,][]{Matsumoto05} and the recent measurement in the optical by the {\it New Horizons} spacecraft \citep{Lauer20}.

An alternative methodology is based on deep galaxy counts, which allow only a local EBL determination \citep[\eg][]{MadauPozzetti00,Dole06,Bethermin11}. This approach is based on data from galaxy surveys, and although robust, can be affected by cosmic variance \citep[\eg][]{somerville04} and limited by survey completeness. Probably, the most complete work to date is \citet{D16}, where the authors compute galaxy counts at 22 bands, from the UV through the FIR, using data from the following surveys: Cosmic Evolution Survey, Galaxy and Mass Assembly, and also Cosmic Assembly Near-Infrared Deep Extragalactic Legacy Survey (CANDELS).

Gamma rays can interact by pair-production with EBL photons in the intergalactic space. Therefore, the observation of high energy extragalactic sources is affected by EBL attenuation \citep[\eg][]{biteau15,dominguez15,paliya19}. This attenuation is dependent on the $\gamma$-ray photon energy and the distance to the source. Several groups, using different approaches, have extracted EBL information from the observation of blazars \citep[\eg][]{aharonian06,aharonian07,mazin07,magic08_3c279,finke09,georgan10,abdo10_EBL,meyer12,ebl12,abdalla17,Fermi18,acciari19,Abeysekara19,Desai19} and $\gamma$-ray bursts \citep{abdo10_EBL,Desai17}.

Empirical EBL models can be built using different methodologies that allow an estimate of the evolving EBL SED. First, purely from fitting optical depths obtained from $\gamma$-ray observations and constructing galaxy emissions that are compatible with the measured attenuation \citep{Fermi18}. Second, one can use semianalytical models of galaxy formation \citep[\eg][]{Somerville12,Gilmore12,inoue13}. These approaches link stellar evolution models with hierarchical scenarios based on the formation of cold dark matter halos to derive the CSFH. 

Third, some models are based on converting the observed CSFH to galaxy emissions using some prescriptions \citep[\eg][]{Kneiske10, F10}. For instance, \citet{Andrews18} use the CSFH modelled by \citet{Driver13} and calculate the total luminosity density up to $z = 4$. In \citet{Khaire19}, they complement some of the most up-to-date CSFH measurements \citep{MD14, HaartMadau12, Behroozi13} with some quasar emissivity data, the H I gas distribution in the IGM from simulations \citep{Inoue14} and the ionizing escape fraction evolution with redshift \citep{Khaire17}. Finally, they compute the overall local diffuse cosmic background from TeV to FIR energies, accounting for IGM attenuation.

Fourth, there are models based on galaxy luminosity functions, or galaxy luminosity densities, and some combination of galaxy SEDs \citep[\eg][]{malkan01,Fr08, D11, H12, Driver15,stecker16, Fr17, Andrews17, Driver18, Bellstedt20, Malkan20}. In particular, our previous model \citep{D11} constructs the EBL SED using a galaxy sample of approximately 6,000 galaxies observed up to $z\sim 1$ selected from the All–wavelength Extended Groth Strip International Survey \citep[AEGIS,][]{davis07} on the Extended Groth Strip field. The multiwavelength SED of these galaxies, which have their reddest detection at $24~\mu$m, were convolved with a $K$-band galaxy luminosity function measured up to $z\sim 4$. 

Here, we present a new determination of the evolving EBL-SED purely based on galaxy data. This original approach uses multiwavelength observations of more than 150,000 galaxies on the five different fields of CANDELS, thus reducing uncertainties related to cosmic variance. This sample is a combination of the deepest galaxy surveys in UV (e.g., with GALEX), optical and NIR (with the {\it Hubble Space Telescope} and ground-based telescopes), MIR (with the \emph{Spitzer Infrared Space Telescope}), FIR (with \emph{Spitzer} and the  {\it Herschel Space Observatory}) and contains a significant number of galaxies detected up to $z\sim 6$.


Our goal is reducing the uncertainties of current EBL models, which are mainly from (1) the EBL-SED evolution, and (2) the far-IR region. We approach (1) recovering a complete EBL evolution up to $z\sim 6$, and a better estimation of the FIR coverage (2) is given by accounting for the emissivity of non-detected sources in the IR. There is an accompanying paper focused on the derived optical depth and the implications for the propagation of $\gamma$ rays in space (Dom\'inguez et al., in preparation).

This paper is organized as follows. In Sect.\  \ref{sec:data} we describe our galaxy sample based on the CANDELS surveys and ancillary data, whereas Sect.\ \ref{sec:method} explains the methodology and our EBL model formalism. In Sect.\ \ref{sec:results} we show the results and discuss them. Finally, a summary is presented in Sect.\ \ref{sec:conclusions}.

Throughout this analysis, a \citet{Chabrier2003} initial mass function (IMF) and a standard $\Lambda$CDM cosmology is used, with a matter density parameter $\Omega_{M}$ = 0.3, a vacuum energy density parameter $\Omega_{\Lambda}$ = 0.7, and Hubble constant $H_0$ = 70~km s$^{-1}$ Mpc$^{-1}$.

%% file: 2_data.tex
\begin{table}
\caption{Central equatorial coordinates, sky-projected area and number of identified sources of the 5 CANDELS fields.}

\begin{tabular}{lcccc}
\hline
Field Name & RA (J2000) & DEC (J2000) & Area$^1$ & No. of sources$^2$ \\
\hline
GOODS--S & 03:32:29.97 & $-$ 27:47:43.62 & 173 & 34~930 \\
UDS & 02:17:36.22 & $-$ 05:12:02.61 & 172 & 35~932 \\
COSMOS & 10:00:30.55 & + 02:21:29.60 & 216 & 38~671 \\
EGS & 14:19:48.38 & + 52:54:04.53 & 206 & 41~457 \\
GOODS--N & 12:36:54.50 & + 62:14:11.01 & 173 & 35~445 \\
\hline
\end{tabular}
$^1$ Units: arcmin$^2$\\
$^2$ Sources included in the original catalogs\\
\label{tab:candels_info}
\end{table}

\begin{figure*}
\centering
    \centering
    \includegraphics[width=0.80\textwidth]{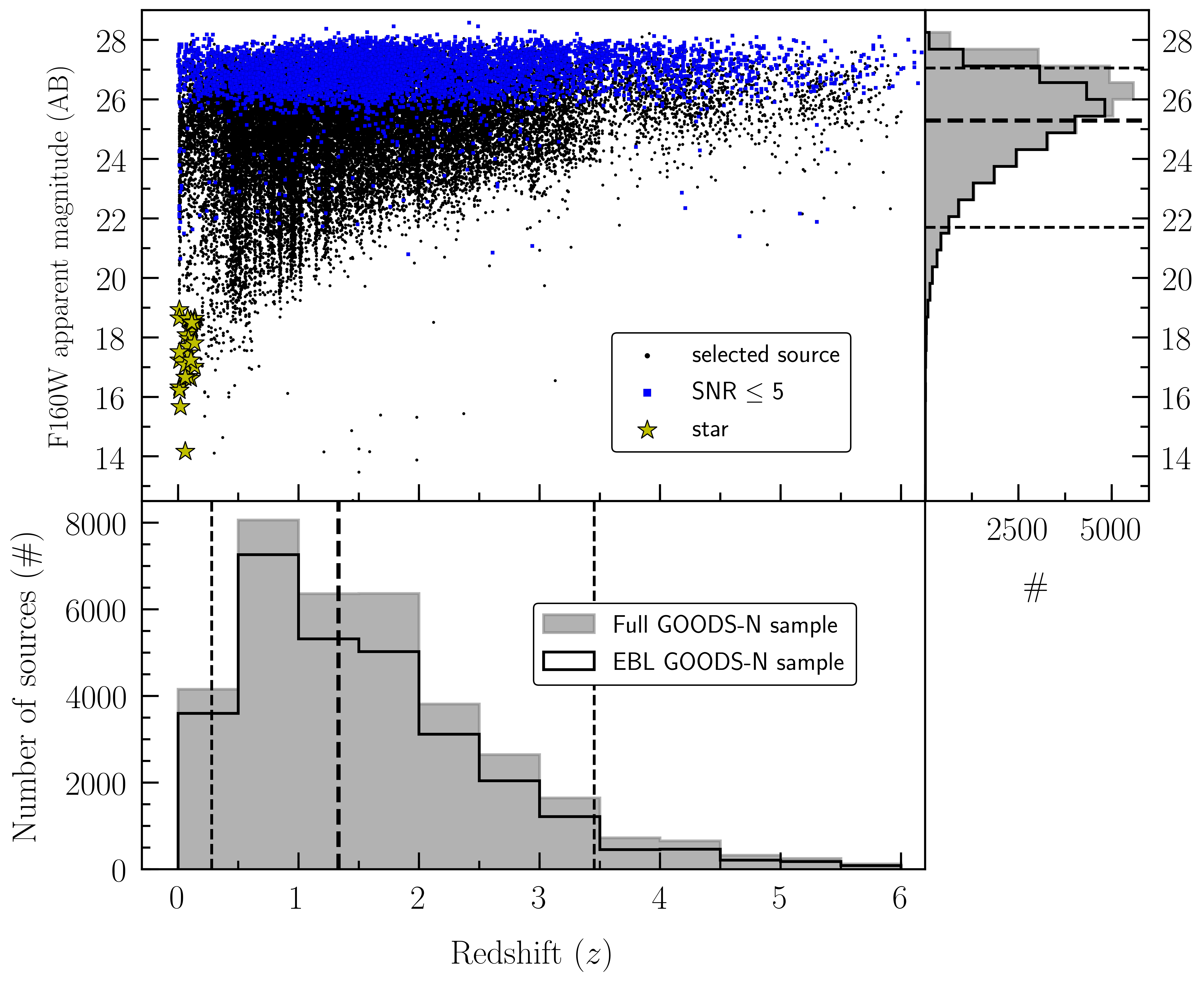}
\caption{Example of F160W magnitude versus redshift diagram showing the filtering process followed for the GOODS--N field. Black dots indicates final selected galaxies while blue points and stars are excluded sources due to S/N $\leq 5$ and ($z \leq 0.15$)$~\&~$(m$_{F160W} \leq 19$) criteria. Bottom and top-left panels show histograms for the same variables before and after the selection criteria was applied. Median, 0.05 and 0.95 quantiles values are indicated by black dotted lines for both the GOODS--N F160W magnitude (\autoref{tab:filtering}) and redshift distributions.}
\label{fig:goodsn_filtering}
\end{figure*}

\begin{figure}
\centering
    \centering
    \includegraphics[width=0.90\columnwidth]{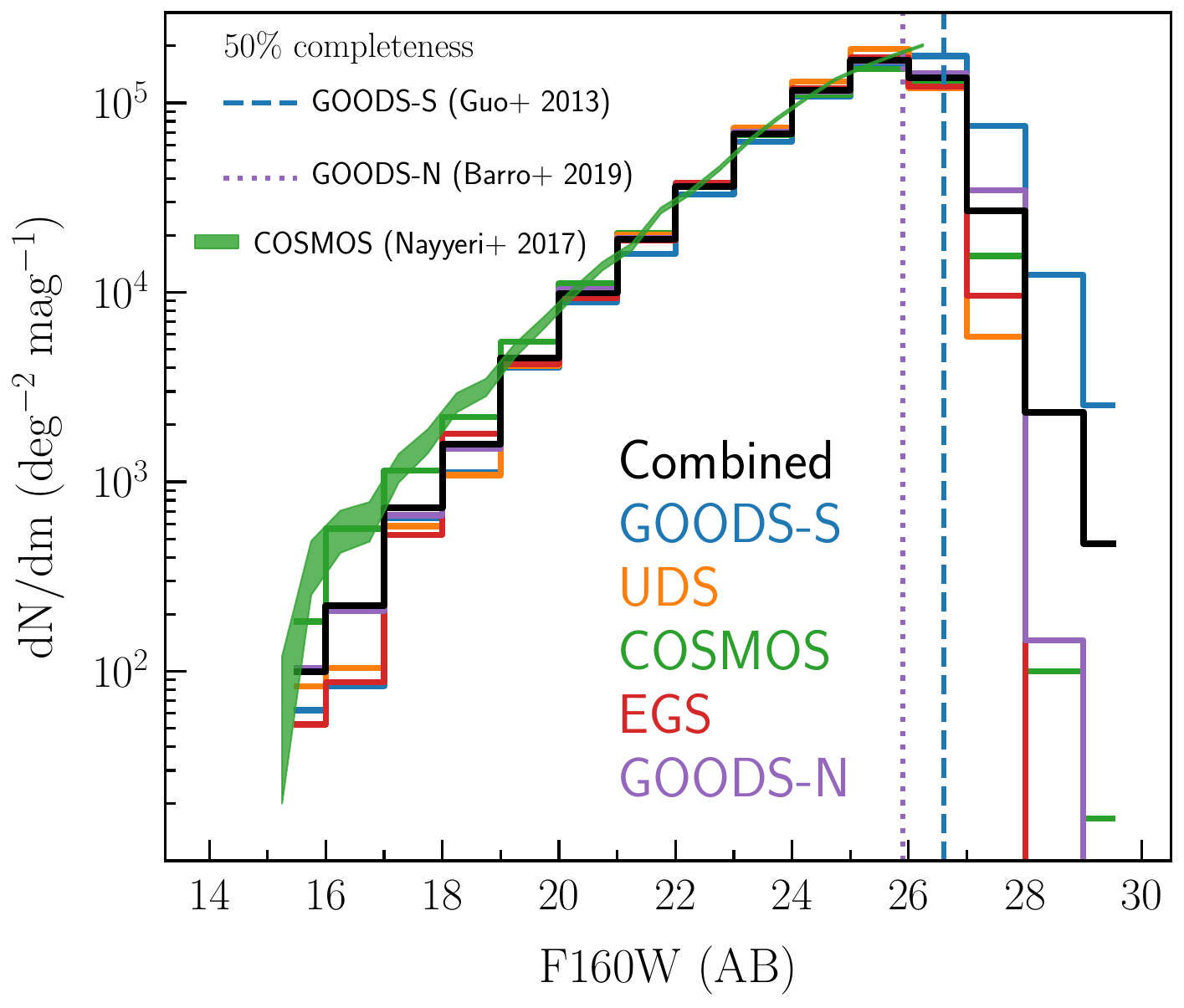}
\caption{Galaxy number counts on every CANDELS field for our final F160W selected sample. Vertical dashed lines limit the F160W(AB) 50$\%$ level of completeness calculated by \citet{Guo13} for GOODS--S and by \citetalias{Barro19} for GOODS--N, for comparison. Red shadowed region represents the number counts calculations by \citet{Nayyeri17} for the COSMOS field.}
\label{fig:ncounts}
\end{figure}

\begin{figure*}
    \includegraphics[height=4.98cm]{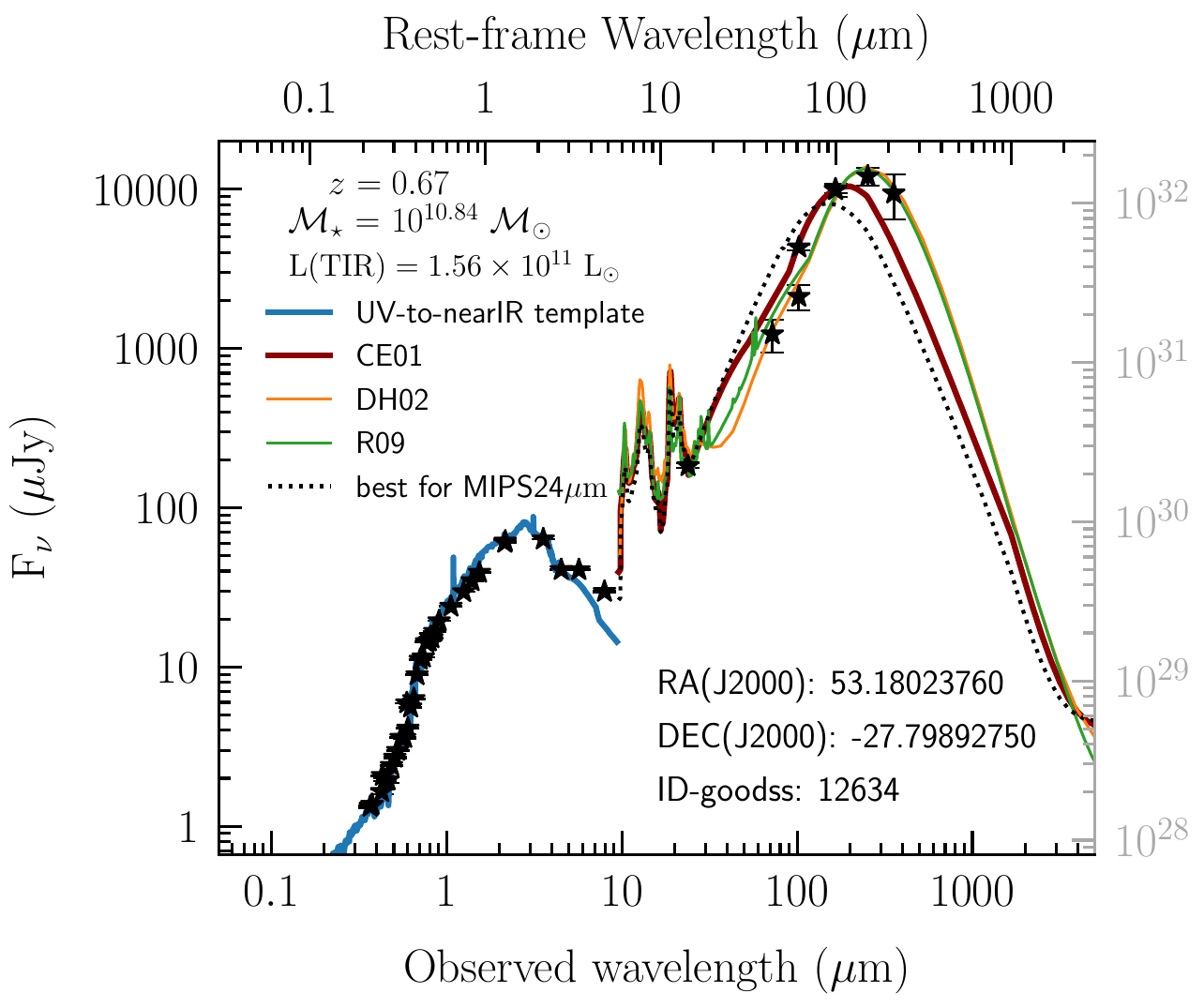}
    \includegraphics[height=4.98cm]{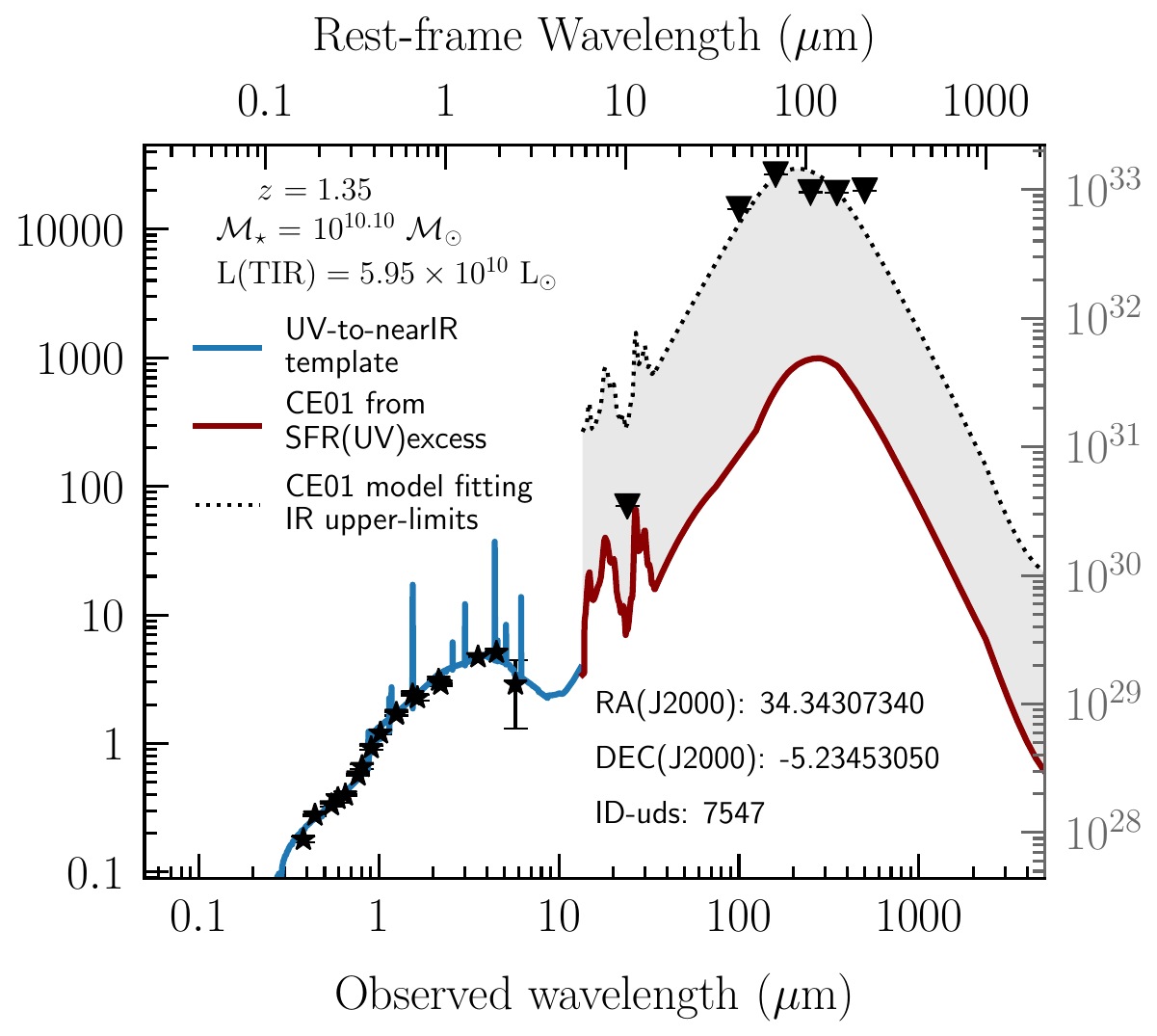}
    \includegraphics[height=4.98cm]{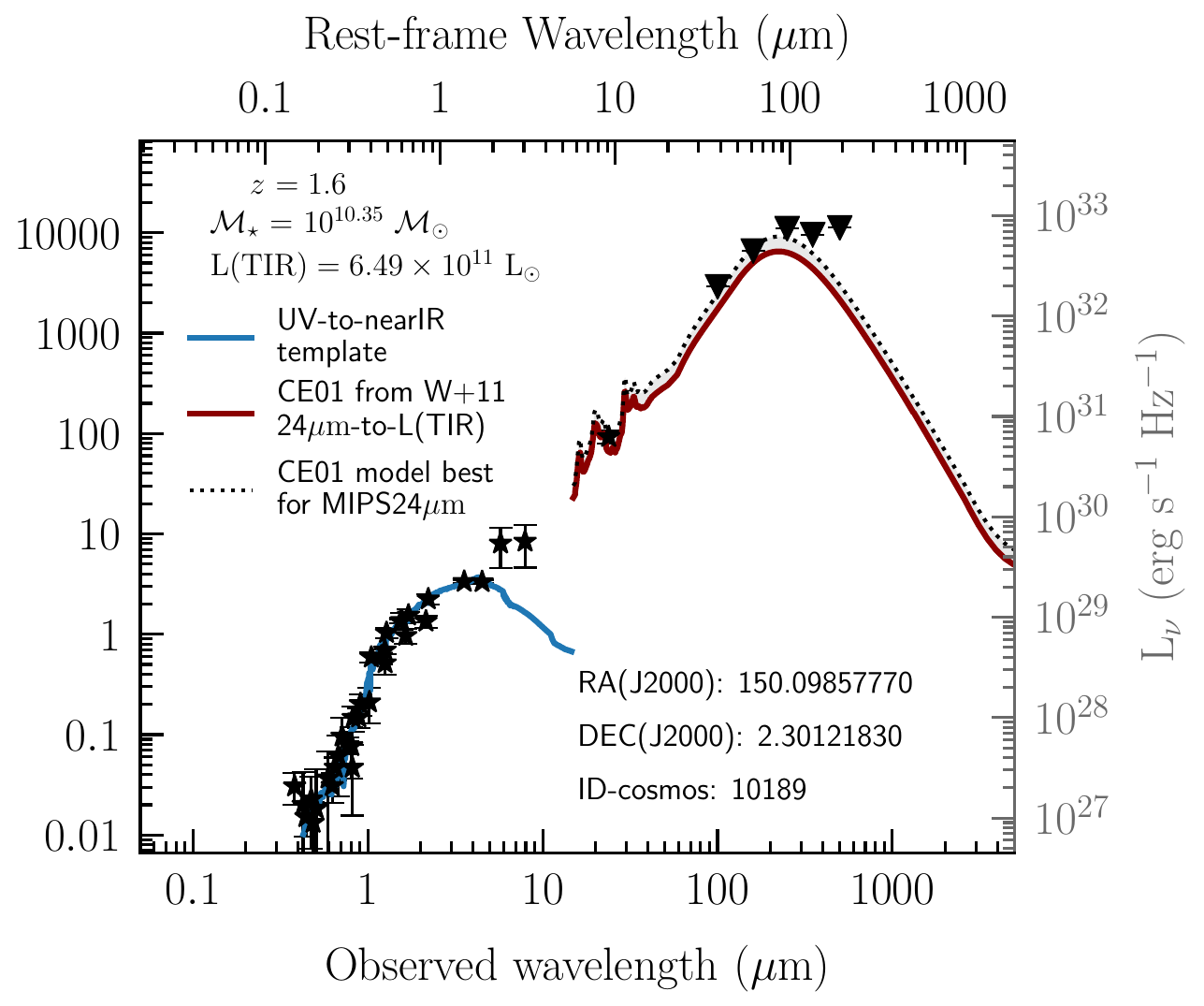}
\caption{Examples of three reconstructed galaxy--SEDs for the 3 different types of source presented in Sect.\ \ref{sec:method} and sorted by redshift (see text). \emph{Left}: galaxy with combined MIPS+Herschel FIR detections; the red, orange and green solid lines are the associated \citetalias{ChEl01}, \citetalias{DaHe02} and \citetalias{Rieke09} dust emission models (the dotted line fits only \citetalias{ChEl01} templates to MIPS/24~$\mu$m point). \emph{Middle}: galaxy with no-IR counterpart by MIPS or Herschel; the solid red line comes from our L(TIR) using the SFR(UV)-excess, while the dashed line is the best fit with upper-limits. \emph{Right}: galaxy with only MIPS/24~$\mu$m detection in the FIR; again, the solid red line correspond to our associated \citetalias{ChEl01} template following the 24$~\mu$m-to-L(TIR) \citetalias{Wuyts11} relation, while the dotted line is the best fit considering also Herschel upper-limits. The blue curve corresponds to the associated \citetalias{Barro19} stellar emission UV-to-NIR template for all the cases. CANDELS identifier, equatorial coordinates (RA, DEC), redshift, stellar mass, and measure/inferred L(TIR) are shown for every object.}
\label{fig:seds}
\end{figure*}

\begin{figure}
\centering
    \centering
    \includegraphics[width=0.90\columnwidth]{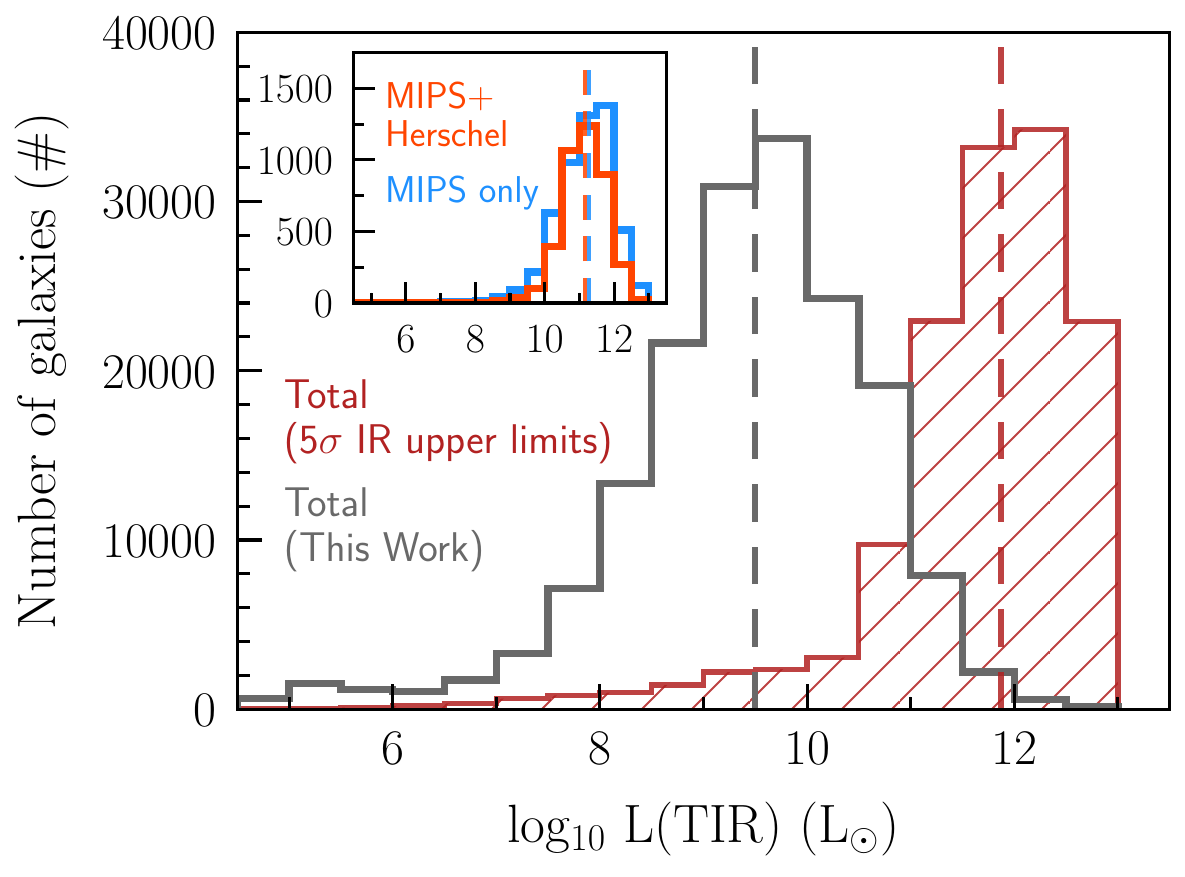}
\caption{L(TIR) number distribution when assuming FIR upper-limits for those sources with no complete IR detection (red dashed histogram) and after our own L(TIR) estimations (grey). In the inset, the number of sources with measured L(TIR) due to combined MIPS+Herchel or MIPS-only detections (\citetalias{Wuyts11} relation applied) are kept in orange and blue, respectively. The median of the every distribution is plotted with a dashed vertical line.}
\label{fig:ltir_hist}
\end{figure}

\begin{figure*}
\centering
    \centering
    \includegraphics[width=0.85\textwidth]{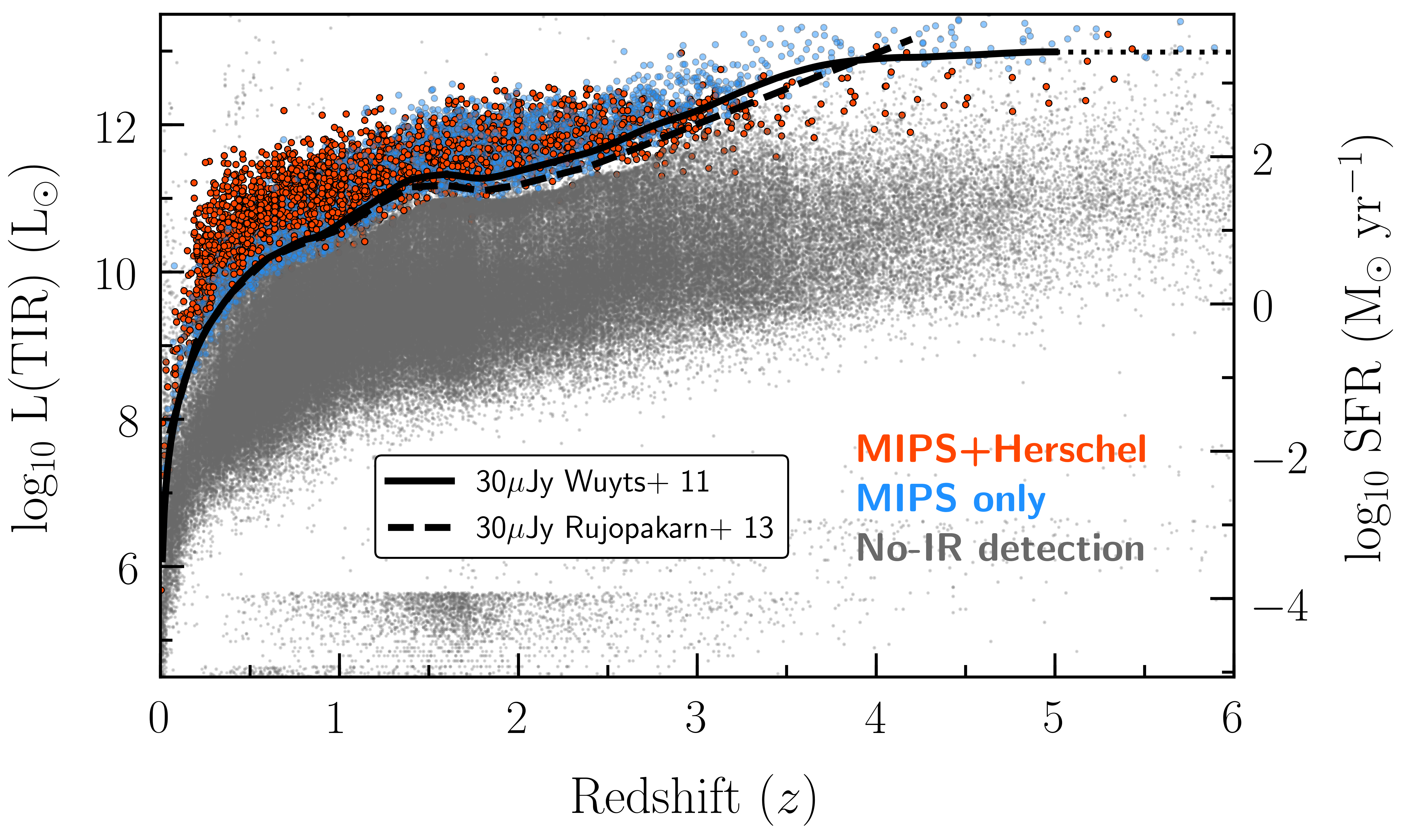}
\caption{L(TIR) versus redshift distribution. Orange points represent those sources with MIPS+Herschel combined detections (4~071) and whose L(TIR) have been directly measured. Blue points are galaxies with only MIPS/24~$\mu$m detection (5~391), and their corresponding L(TIR) comes from applying the \citetalias{Wuyts11} relation. Finally, grey points are sources no detected in the IR (144~985), and our SFR(UV)-excess recipe was applied to obtain the L(TIR). For comparison, the GOODS--S/N 30~$\mu$Jy limiting fluxes curves are plotted using the \citetalias{Wuyts11} and \citet{R13} expressions (\citetalias{Wuyts11} extrapolated out to $z=6$).}.
\label{fig:ltir_z}
\end{figure*}

\begin{figure*}
    \includegraphics[height=8cm]{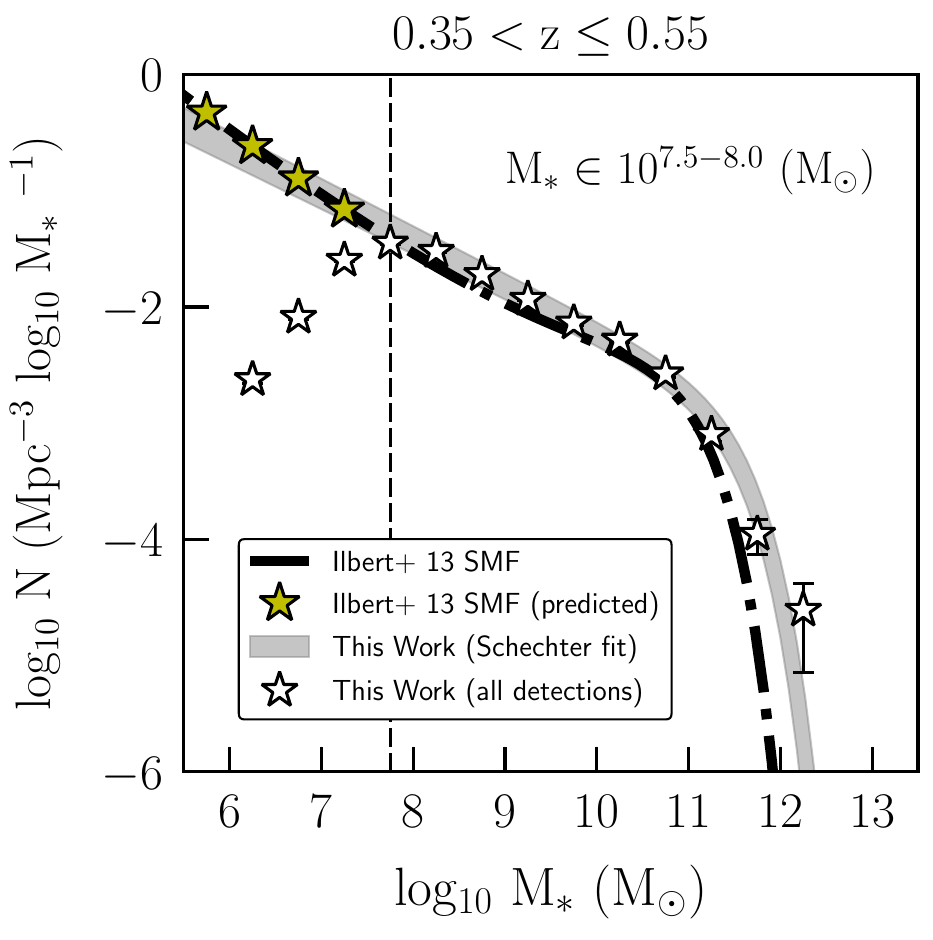}
    \hspace{0.5cm}
    \includegraphics[height=8cm]{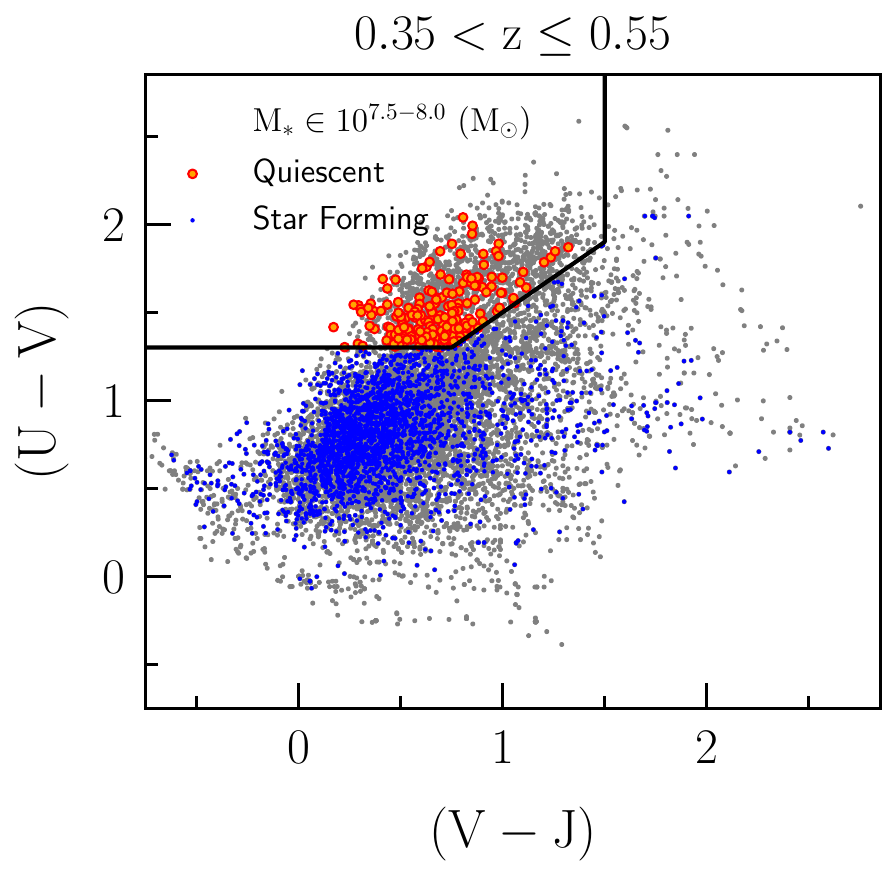}
\caption{Our completeness correction sketch. \emph{Left}: the observed mass-density of objects for our EBL sample at $0.35 \leq z < 0.55$, in white open stars; the corresponding mass completeness limit is pointed by the dashed vertial line: $M_*^{lim}=10^{7.5-8.0}$ ~$M_{\odot}$. On the other hand, the prediction from the \citet{Ilbert13} SMF for galaxies above this mass limit is highlighted by yellow stars. The grey shaded area represent the adopted SMF and our own Schechter fit to the data (1$\sigma$ band uncertainty). \emph{Right}: UVJ color-color selection diagram \citep{Williams09} at $0.35 \leq z < 0.55$. The orange and blue circles indicate the quiescent and SF galaxies selected by this method on the corresponding mass completeness limit bin.}.
\label{fig:smf}
\end{figure*}

\section{Data}\label{sec:data}
\subsection{The CANDELS multi-wavelength survey}\label{sub:data_description}
CANDELS\footnote{\url{http://arcoiris.ucolick.org/candels/}} is one of the most ambitious {\it Hubble Space Telescope} (HST) observational programs \citep{Grogin11, Koekmoer11}. It involves more than 4 months of cumulative observing time between 2010 and 2013. The CANDELS HST surveys have selected a set of around 200~000 galaxies up to $z = 8$ using two of its instruments: the Advance Camera for Surveys (ACS) and the Wide Field Camera 3 (WFC3). Both provide NUV to NIR data collection (0.435-1.6~$\mu$m) thanks to broadband filter imaging. The CANDELS selection criteria enforced observation in WFC3/F160W filter (1.6~$\mu$m central wavelength, $H$-band) focusing on five fields: (1) Great Observatories Origins Deep Survey South (GOODS--S), (2) Ultra Deep Survey (UDS), (3) Cosmic Evolution Survey (COSMOS), (4) Extended Groth Strip (EGS), and (5) Great Observatories Origins Deep Survey North (GOODS--N), sorted by date of publication.

The CANDELS observing program is composed by two observational modes with different $H$-band limiting apparent magnitude (m$_H^{lim}$): CANDELS/Wide (m$_H^{lim} \sim$ 27 AB) and CANDELS/Deep (m$_H^{lim} \sim$ 27.7 AB). All these fields have a sky-projected area of around 200~arcmin$^2$ and host $\sim$35~000 detected sources each. \autoref{tab:candels_info} shows a detailed description of the central RA/DEC coordinates, covered area, and number of sources. For this work, we use the catalogs published by the CANDELS team and released through the Rainbow database\footnote{The Rainbow database is operated by the Centro de Astrobiología (CAB/INTA), partnered with the University of California Observatories at Santa Cruz (UCO/Lick,UCSC)}.

\begin{table}
\caption{Flux upper-limits for CANDELS IR bands.}
\begin{tabular}{l|ccccc}
\hline
& \scriptsize{GOODS-S} & \scriptsize{UDS} & \scriptsize{COSMOS} & \scriptsize{EGS} & \scriptsize{GOODS-N} \\
\hline
MIPS24~($\mu$Jy) & 30 & 70 & 70 & 45 & 30 \\
PACS100~(mJy) & 1.1 & 14.4 & 2.9 & 8.7 & 1.6 \\
PACS160~(mJy) & 3.4 & 26.7 & 6.6 & 13.1 & 3.6 \\
SPIRE250~(mJy) & 8.3 & 19.4 & 11.0 & 14.7 & 9.0 \\
SPIRE350~(mJy) & 11.5 & 19.2 & 9.6 & 17.3 & 12.9 \\
SPIRE500~(mJy) & 11.3 & 20.0 & 11.2 & 17.9 & 12.6 \\
\hline
\end{tabular}
\label{tab:IRuppLIM}
\end{table}

The CANDELS/GOODS--S sample is composed by 34~930 sources selected via HST/WFC3/F160W detection over HST/WFC3 F105W, F125W, and F160W original exposures on a 173~arcmin$^{2}$ sky-area \citep{Guo13}. The mosaic reaches a 5$\sigma$ limiting depth (within an aperture radius of 0.17~arcsec) of 27.4~AB and 28.2~AB for CANDELS Wide and Deep modes. In addition to WFC3 bands, the catalog also includes data from UV ($U$-band from both CTIO/MOSAIC and VLT/VIMOS), optical (HST/ACS F435W, F606W, F775W, F814W, and F850LP), and NIR (HST/WFC3 F098M, VLT/ISAAC K$_s$, VLT/HAWK-I K$_s$) observations. The Hubble Deep field UV Legacy Survey (HDUV) provides complementary observations to the CANDELS/GOODS--S Deep mode over a total area of $\sim$100~arcmin$^2$ in the two filters F275W and F336W \citep{Oesch18}.

The CANDELS/UKIDSS-UDS sample comprises a 173~arcmin$^{2}$ area with 35~932 HST WFC3/F125W and F160W selected sources \citep[5$\sigma$ limiting depth of 27.5~AB,][]{Galametz13}. This includes observations in the visible through HST/ACS (F606W and F814W), u-band data from CFHT/Megacam, B, V, R$_c$, i', and z' band data from Subaru/Suprime-Cam, Y and K$_s$ band data from VLT/HAWK-I, and also J, H, and K band data from UKIDSS (Data Release 8, UKIRT/WFCAM).

The CANDELS/COSMOS catalog is based on HST/WFC3/F160W detection selection criteria with a 27.6~AB limiting magnitude (5$\sigma$ depth) over a 216~arcmin$^{2}$ field, resulting in 38~671 sources with photometric data in several bands from UV to the IR \citep{Nayyeri17}. This includes broadband photometry from HST/ACS (F606W, F814W), HST/WFC3 (F125W, F160W), CFHT (MegaPrime/$ugriz$), Subaru (Suprime-Cam/ B, $griz$ $^+$, V) and VISTA (VIRCAM/ Y, J, H, K$_{s}$) in the visible and NIR bands, along with intermediate and narrow band photometry from Subaru and mediumband data from Mayall/NEWFIRM (J1, J2, J3, Y1, Y2, K), respectively.

The photometric catalog in the CANDELS/EGS is built on the HST/ACS and WFC observations and selected 41~457 sources reaching a F160W 5$\sigma$ depth of 26.6~AB, over an area of 206~arcmin$^{2}$ \citep{Stefanon17}. This sample includes the following bands: HST/ACS F606W and F814W; HST/WFC3 F125W, F140W, and F160W; CFHT/Megacam u*,g',r',i' and z'; CFHT/WIRCAM J, H and K$_s$ and Mayall/NEWFIRM J1, J2, J3, H1, H2 and K.

Finally, the F160W selected CANDELS/GOODS-N sources comprise a 173~arcmin$^{2}$ sky-area and 35~445 sources \citep[hereafter \citetalias{Barro19}]{Barro19}. The 5$\sigma$ detection limits (0.17~arcsec aperture radius) of the mosaic ranges between m$_H^\mathrm{lim}$ = 27.8 and 28.7~AB in the Wide and Deep regions. The multi-wavelength photometry has broadband data from the UV ($U$-band from KPNO and LBT), optical (HST/ACS F435W, F606W, F775W, F814W, and F850LP) and near-to-mid IR (HST/WFC3 F105W, F125W, F140W, and F160W; Subaru/MOIRCS K$_s$ and CFHT/Megacam K) observations. In addition to the WFC3, this release also includes data from the 25 medium-band filter observations from the with the Survey for High-$z$ Absorption Red and Dead Sources (SHARDS) program using the Gran Telescopio Canarias \citep{PG13}.

Additionally in all fields, the {\it Spitzer Infrared Space Telescope} provides IRAC/ 3.6, 4.5, 5.8, 8.0~$\mu$m and MIPS/24 and 70~$\mu$m photometry; while Herschel\footnote{Herschel is an ESA space observatory with science instruments provided by European-led Principal Investigator consortia and with important participation from NASA.} does through PACS/ 100 and 160~$\mu$m, SPIRE/ 250, 350 and 500~$\mu$m. A description of the Spitzer and Herschel catalogs and procedures to identify the counterparts within the CANDELS catalogs is presented in \citetalias{Barro19} --see also \citet{Rawle16}--, and the 5$\sigma$ limiting fluxes for Spitzer/MIPS and Herschel/PACS and SPIRE are summarized in \autoref{tab:IRuppLIM}. Also included in the GOODS–S/N fields are the GALEX FUV and NUV photometry bands (0.1530 and 0.2310~$\mu$m, \citet{Morrissey07}). 

A summary of the CANDELS Photometry Data Set listing most of the photometric bands publicly available for the CANDELS mission is presented in \autoref{tab:candels_bands}.

\begin{table*}
\caption{Summary of photometric broad- and intermediate-band data set available in \emph{Rainbow Navigator} for the 5 CANDELS fields.}
\begin{tabular}{lccccccc}
\hline
Instrument & Filter & $\mathrm{\lambda}$ (\AA) & GOODS--S & UDS & COSMOS & EGS & GOODS--N \\
\hline
GALEX & FUV & 1~538 & \checkmark & -- & -- & -- & \checkmark \\
GALEX & NUV & 2~315 & \checkmark & -- & -- & -- & \checkmark \\
Blanco/MOSAIC--II & U\_CTIO & 3~734 & \checkmark & -- & -- & -- & -- \\
VLT/VIMOS & U\_VIMOS & 3~722 & \checkmark & -- & -- & -- & -- \\
KPNO4m/Mosaic & U & 3~592 & -- & -- & -- & -- & \checkmark \\
LBT/LBC & U$^{/}$ & 3~633 & -- & -- & -- & -- & \checkmark \\
CFHT/MegaPrime & u & 3~817 & -- & -- & \checkmark & -- & -- \\
CFHT/MegaPrime & g & 4~317 & -- & -- & \checkmark & -- & -- \\
CFHT/MegaPrime & r & 6~220 & -- & -- & \checkmark & -- & -- \\
CFHT/MegaPrime & i & 7~606 & -- & -- & \checkmark & -- & -- \\
CFHT/MegaPrime & z & 8~816 & -- & -- & \checkmark & -- & -- \\
CFHT/MegaCam & u$^{*}$ & 3~860 & -- & \checkmark & -- & \checkmark & -- \\
CFHT/MegaCam & g$^{/}$ & 4~890 & -- & -- & -- & \checkmark & -- \\
CFHT/MegaCam & r$^{/}$ & 6~250 & -- & -- & -- & \checkmark & -- \\
CFHT/MegaCam & i$^{/}$ & 7~690 & -- & -- & -- & \checkmark & -- \\
CFHT/MegaCam & z$^{/}$ & 8~880 & -- & -- & -- & \checkmark & -- \\
CFHT/MegaCam & K & 21~347 & -- & -- & -- & -- & \checkmark \\
Subaru/Suprime--Cam & B & 4~448 & -- & \checkmark & \checkmark & -- & -- \\
Subaru/Suprime--Cam & g$^{+}$ & 4~761 & -- & -- & \checkmark & -- & -- \\
Subaru/Suprime--Cam & V & 5~470 & -- & \checkmark & \checkmark & -- & -- \\
Subaru/Suprime--Cam & r$^{+}$ & 6~276 & -- & -- & \checkmark & -- & -- \\
Subaru/Suprime--Cam & R$_c$ & 6~500 & -- & \checkmark & -- & -- & -- \\
Subaru/Suprime--Cam & i$^{+}$ & 7~671 & -- & \checkmark & \checkmark & -- & -- \\
Subaru/Suprime--Cam & z$^{+}$ & 9~096 & -- & \checkmark & \checkmark & -- & -- \\
Subaru/MOIRCS & K$_{s}$ & 21~577 & -- & -- & -- & -- & \checkmark \\
HST/ACS & F435W & 4~317 & \checkmark & -- & -- & -- & \checkmark \\
HST/ACS & F606W & 5~918 & \checkmark & \checkmark & \checkmark & \checkmark & \checkmark \\
HST/ACS & F775W & 7~693 & \checkmark & -- & -- & -- & \checkmark \\
HST/ACS & F814W & 8~047 & \checkmark & \checkmark & \checkmark & \checkmark & \checkmark \\
HST/ACS & F850LP & 9~055 & \checkmark & -- & -- & -- & \checkmark \\
HST/WFC3 & F098M & 9~851 & \checkmark & -- & -- & -- & -- \\
HST/WFC3 & F105W & 10~550 & \checkmark & -- & -- & -- & \checkmark \\
HST/WFC3 & F125W & 12~486 & \checkmark & \checkmark & \checkmark & \checkmark & \checkmark \\
HST/WFC3 & F140W & 13~970 & -- & -- & -- & \checkmark & \checkmark \\
HST/WFC3 & F160W & 15~370 & \checkmark & \checkmark & \checkmark & \checkmark & \checkmark \\
VISTA/VIRCAM & Y & 10~210 & -- & -- & \checkmark & -- & -- \\
VISTA/VIRCAM & J & 12~524 & -- & -- & \checkmark & -- & -- \\
VISTA/VIRCAM & H & 16~431 & -- & -- & \checkmark & -- & -- \\
VISTA/VIRCAM & Ks & 21~521 & -- & -- & \checkmark & -- & -- \\
VLT/ISAAC & Ks & 21~605 & \checkmark & -- & -- & -- & -- \\
UKIRT/WFCAM & J & 12~510 & -- & \checkmark & -- & -- & -- \\
UKIRT/WFCAM & H & 16~360 & -- & \checkmark & -- & -- & -- \\
UKIRT/WFCAM & K & 22~060 & -- & \checkmark & -- & -- & -- \\
Mayall/NEWFIRM & J1 & 10~470 & -- & -- & \checkmark & \checkmark & -- \\
Mayall/NEWFIRM & J2 & 11~950 & -- & -- & \checkmark & \checkmark & -- \\
Mayall/NEWFIRM & J3 & 12~790 & -- & -- & \checkmark & \checkmark & -- \\
Mayall/NEWFIRM & H1 & 15~610 & -- & -- & \checkmark & \checkmark & -- \\
Mayall/NEWFIRM & H2 & 17~070 & -- & -- & \checkmark & \checkmark & -- \\
Mayall/NEWFIRM & K & 21~700 & -- & -- & \checkmark & \checkmark & -- \\
CHFT/WIRCAM & J & 12~540 & -- & -- & -- & \checkmark & -- \\
CHFT/WIRCAM & H & 16~360 & -- & -- & -- & \checkmark & -- \\
CHFT/WIRCAM & Ks & 21~590 & -- & -- & -- & \checkmark & -- \\
VLT/HAWK--1 & Y & 10~190 & -- & \checkmark & -- & -- & -- \\
VLT/HAWK--1 & Ks & 21~463 & \checkmark & \checkmark & -- & -- & -- \\
Spitzer/IRAC & 3.6~$\mu$m & 35~508 & \checkmark & \checkmark & \checkmark & \checkmark & \checkmark \\
Spitzer/IRAC & 4.5~$\mu$m & 44~960 & \checkmark & \checkmark & \checkmark & \checkmark & \checkmark \\
Spitzer/IRAC & 5.8~$\mu$m & 57~245 & \checkmark & \checkmark & \checkmark & \checkmark & \checkmark \\
Spitzer/IRAC & 8.0~$\mu$m & 78~840 & \checkmark & \checkmark & \checkmark & \checkmark & \checkmark \\
Spitzer/MIPS & MIPS/24 & 238~440 & \checkmark & \checkmark & \checkmark & \checkmark & \checkmark \\
Spitzer/MIPS & MIPS/70 & 724~936 & \checkmark & \checkmark & \checkmark & \checkmark & \checkmark \\
Herschel/PACS & PACS/100 & 1~022~522 & \checkmark & \checkmark & \checkmark & \checkmark & \checkmark \\
Herschel/PACS & PACS/160 & 1~656~067 & \checkmark & \checkmark & \checkmark & \checkmark & \checkmark \\
Herschel/SPIRE & SPIRE/250 & 2~531~298 & \checkmark & \checkmark & \checkmark & \checkmark & \checkmark \\
Herschel/SPIRE & SPIRE/350 & 3~558~741 & \checkmark & \checkmark & \checkmark & \checkmark & \checkmark \\
Herschel/SPIRE & SPIRE/500 & 5~111~803 & \checkmark & \checkmark & \checkmark & \checkmark & \checkmark \\
\hline
\end{tabular}
\label{tab:candels_bands}
\end{table*}

\begin{table*}
\caption{Data filtering procedure for the CANDELS EBL sample and resulting median ($\mathcal{H}$), 0.05 (05$\%$) and 0.95 (95$\%$) quantiles values for every F160W (AB) magnitude distribution. The filtered sample is composed by 154,447/186,435 sources ($82.8\%$ of the total).}
\begin{tabular}{lcccccc}
\hline
Cut & GOODS-S & UDS & COSMOS & EGS & GOODS-N & Total\\
\hline
CANDELS & 34~930 & 35~932 & 38~671 & 41~457 & 35~445 & 186~435\\
F160W detection & 34~848 & 35~752 & 38~216 & 41~353 & 35~440 & 185~609\\
S/N $\geq$ 5 & 32~224 & 28~222 & 33~327 & 32~577 & 29~269 & 155~619\\
After removing stars & 32~170 & 28~742 & 33~297 & 32~518 & 29~242 & 155~969\\
0 $\leq$z$<$ 6 & 31~736 & 28~496 & 32~906 & 32~362 & 28~947 & 154~447\\
\hline 
F160W: $\mathcal{H}^{95\%}_{05\%}$ &
25.64$^{27.48}_{22.10}$ &
25.13$^{26.56}_{21.79}$ &
25.13$^{26.84}_{21.40}$ & 25.14$^{26.69}_{21.72}$ & 25.29$^{27.05}_{21.70}$ &
25.25$^{27.00}_{21.72}$\\
\hline
\end{tabular}
\label{tab:filtering}
\end{table*}

\subsection{Sample selection and data filtering}\label{sub:data_filtering}
In order to work with a well-characterized photometric sample, that is, in which every galaxy-SED can be recovered with acceptable uncertainties, we revise the parent CANDELS catalogs and purge problematic sources.

Oue sample selection process is the following:

\begin{enumerate}
    \item F160W detection is required.
    \item F160W signal-to-noise ratio (S/N) should be larger than 5, \ie $\mathrm{S/N}_{F160W} = \dfrac{f_{F160W}}{\Delta f_{F160W}} \geq  5$.
    \item We remove identified stars. Additional sources whose SEDs might be contaminated by the emission from nearby bright stars are purged by an additional cut in the F160W apparent magnitude ($m_{F160W} \leq 19$) for the lowest redshift sources ($z \leq 0.15$). This step was optimized based on the BzK diagrams stellar locus and other point-like sources' parameters provided in the CANDELS catalogs, such as for instance \texttt{CLASS$\_$STAR} \citep[see \eg][]{B11a,B11b,Guo13}.
    \item We select only sources with $0\leq z < 6$ (either photometric or spectroscopic redshift).
\end{enumerate}

In \autoref{tab:filtering}, we show the resulting number of sources after every filtering step independently for the five CANDELS fields and the final sample. The median for the F160W magnitude distributions with its 0.05 and 0.95 quantiles are given. Our final sample, that we call EBL sample, is composed by 154~447 out of a total of 186~435 sources (this is, $82.8\%$ of the total). \autoref{fig:goodsn_filtering} shows the filtering procedure on a magnitude versus redshift diagram for the GOODS-N field as example. 

In \autoref{fig:ncounts}, we plot the differential galaxy number counts (number density) for the final EBL sample, and independently for each CANDELS field. We check the representativeness of our EBL sample compared to the parent sample in terms of this number density of sources \citep[\eg][]{King86}. The procedure is the following: we fit a power law to the differential number density of objects, then typically the sample becomes incomplete when the observed differential number density significantly deviates from the best-fit power law at the faint end. For GOODS-S \citep{Guo13} and GOODS-N \citepalias{Barro19}, their actual differential number densities becomes lower than the best-fit power law by a factor of two at 26.6 and 25.9 AB, which is in agreement with the counts from our EBL sample. The COSMOS galaxy number counts from \citet{Nayyeri17} are also plotted for comparison through the red shadowed band.

In the forthcoming Sect. \ref{sub:result_EBLlocal}, we will approach how our sample selection criteria (mainly, F160W S/N cut) can affect the EBL intensity level.

\newpage
\subsection{The Rainbow database}\label{sub:Rainbow}
To access and retrieve CANDELS data, we use the Rainbow Cosmological Surveys Database \citep{B11a,B11b} through the publicly accessible Rainbow Navigator\footnote{\url{http://arcoirix.cab.inta-csic.es/Rainbow_navigator_public}} web interface. This database collects a vast compilation of photometric and spectroscopic data for the CANDELS fields. Although independent SED fitting results are included in the CANDELS catalog papers, we prefer to use a homogenized SED method based on the Rainbow pipeline (see below). In addition, we use the UV- and IR-derived SFRs and the IR dust emission fits provided by the same Rainbow database pipeline, described for the 5 CANDELS fields in \citetalias{Barro19}.

The Rainbow SED pipeline \citep[see][, and \citetalias{Barro19}]{B11a, B11b} use the \emph{Rainbow} template fitting code developed by \citet{PG05}. Briefly, this program applies a $\chi^2$ minimization algorithm as:

\begin{equation}\label{eq:chi2}
    \chi^2=\sum_{i=1}^N \left( \dfrac{\mathrm{F}_{obs,i} - A \times \mathrm{F}_{temp,i}}{\sigma_i^2}\right)^2,
\end{equation}

\noindent
where $F_{obs,i}$ ($F_{temp,i}$) is the observed (template) flux in the $i$th band, with an observational uncertainty $\sigma_i$ . The SED templates are first converted from rest- to observed-frame using a cosmological factor $\times (1+z)$. The redshift is given by the CANDELS photometric or spectroscopic redshift when available. Uncertainties on the redshifts are not included in the analysis but \citet{D11} already showed that this does not produce a significant effect on the EBL intensities. A template renormalization factor, $A$, is used to scale the observed SED to the associated template profile, and it accounts for the stellar mass and the luminosity distance for every galaxy. The UV-to-NIR region (defined in the 0.1 to 5.7~$\mu$m rest-frame wavelength interval) is modelled by stellar population templates, whereas the IR (from 5.7 to 1000~$\mu$m) is fitted independently by different dust emission models.

The stellar population templates are extracted from a semi-empirical library of 1,876 synthetic SEDs, built from GOODS–S/N IRAC selected sources \citep{PG08}, and some AGN empirical templates from \citet{Polletta07}. Assuming a Chabrier IMF and a Calzetti extinction law \citep{Calzetti01}, the stellar emission of the reference templates is characterized using the \texttt{PEGASE2.0} models \citep{PEGASE1997}. The contribution from emission lines and the nebular continuum emitted by ionized gas are also considered. The models are obtained assuming a single stellar population, driven by an exponential star formation law. As a result, each final template is characterised by four physical parameters: the burst time scale, age, metallicity and dust extinction. 

For the IR spectral region, \citetalias{Barro19} fit the observed IR fluxes to three different sets of dust emission templates: \citet[hereafter \citetalias{ChEl01}]{ChEl01}, \citet[\citetalias{DaHe02}]{DaHe02}, and \citet[\citetalias{Rieke09}]{Rieke09}, assuming 5$\sigma$ upper limits to the corresponding IR band in cases where there is no detection (\autoref{tab:IRuppLIM}).

In this work, we use the CANDELS photometric catalogs and redshifts, which together with the associated SED templates by Rainbow, allow us to reproduce the 0.1-1000~$\mu$m SED of every galaxy in our EBL sample. Failed fits and outliers are removed from the working sample by 3$\sigma$ clipping applied to different bands along the galaxy-SEDs distribution. \autoref{fig:seds} shows an example of a galaxy SED from GOODS-S with detection in all bands. This figure also includes the corresponding UV-to-optical SED template as well as the three dust emission models.

%% file: 3_methodology.tex
\begin{figure*}
    \includegraphics[width=0.475\textwidth]{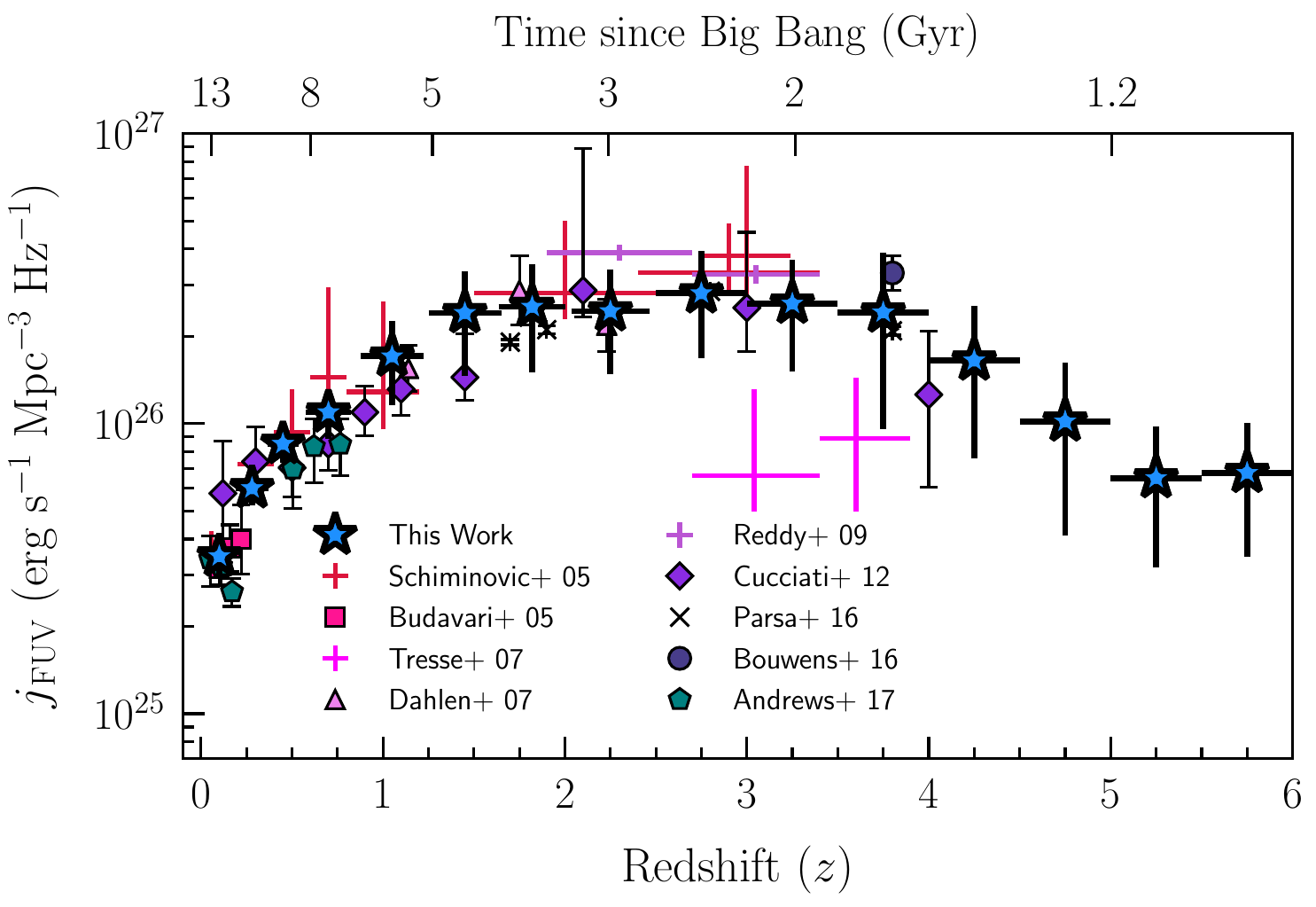}
    \hspace{0.25cm}
    \includegraphics[width=0.475\textwidth]{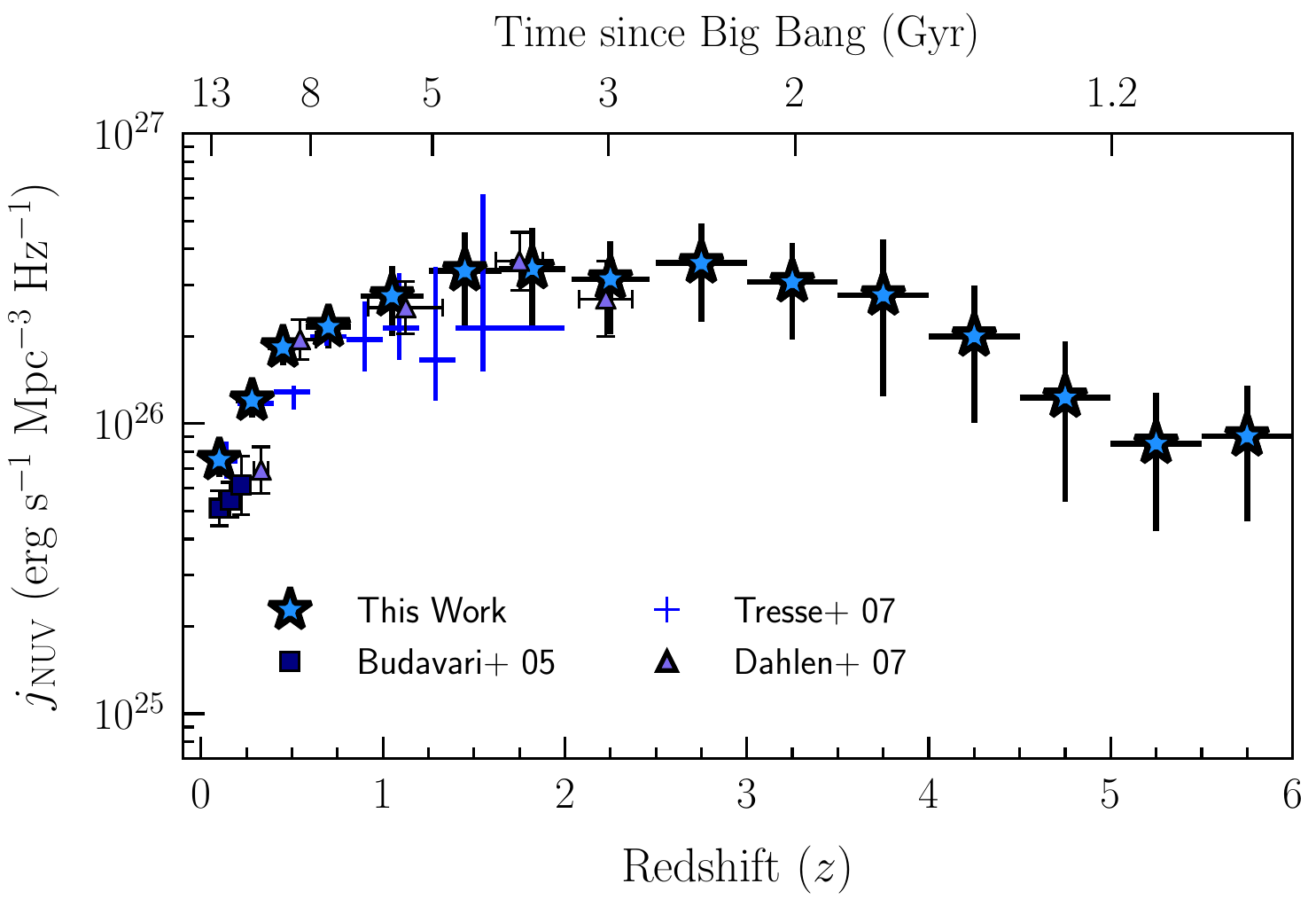}\\

    \vspace{0.4cm}
    \includegraphics[width=0.475\textwidth]{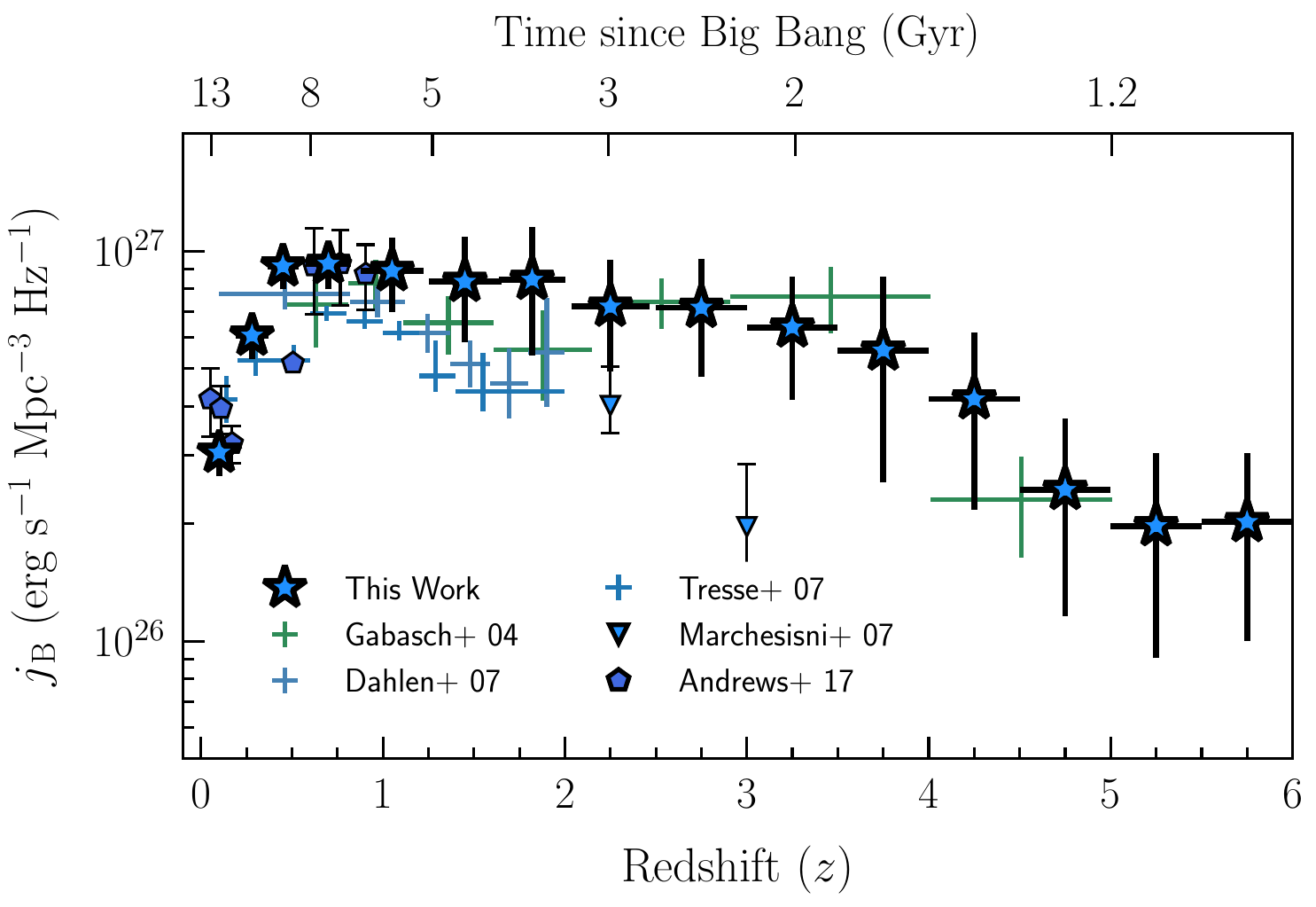}
    \hspace{0.25cm}
    \includegraphics[width=0.475\textwidth]{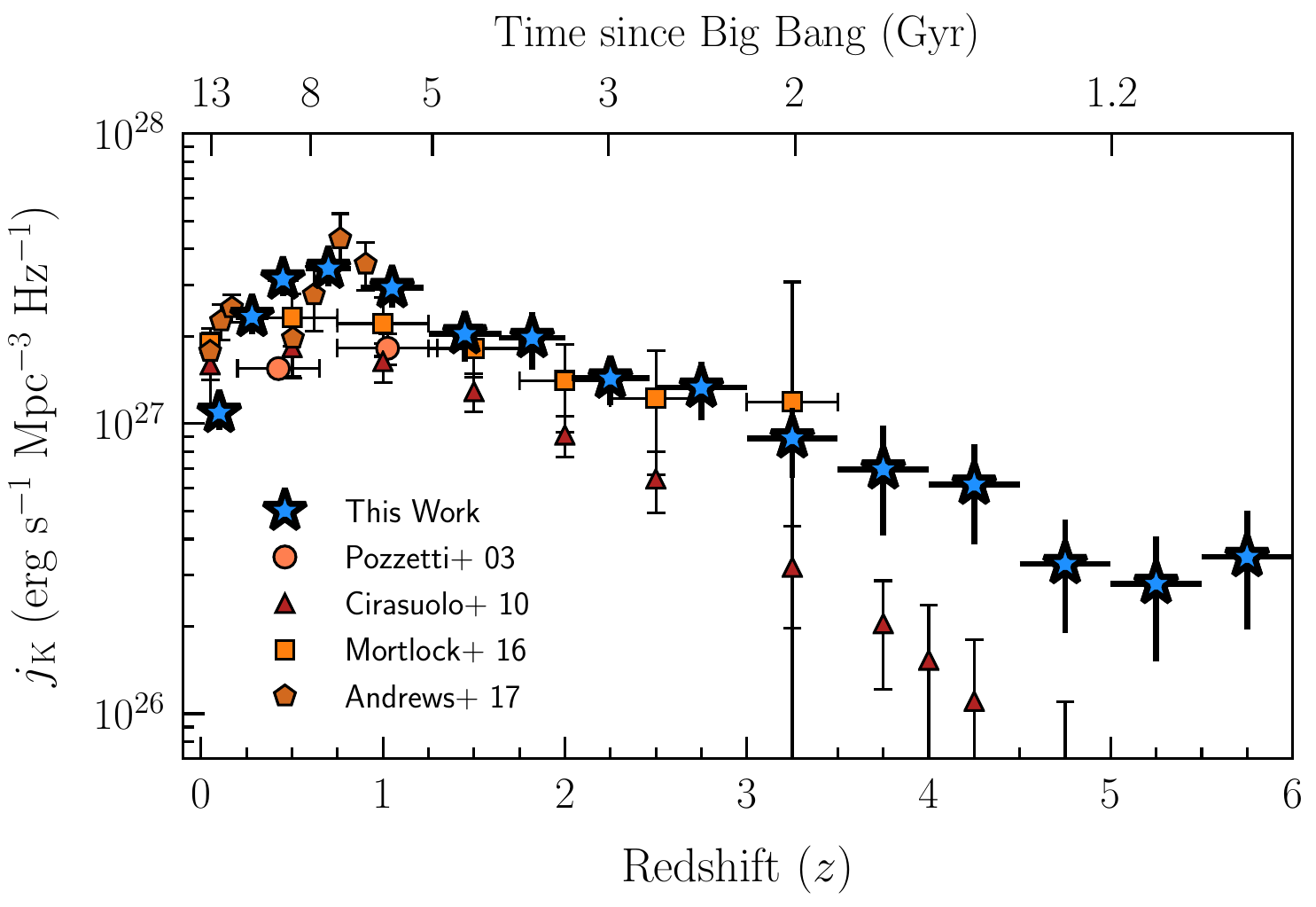}
\caption{Derived FUV (1500~\AA), NUV (2800~\AA), optical ($B$-band, 4400~\AA) and NIR ($K$-band, 2.2$~\mu$m) luminosity densities between $z=0$ and $z=6$ (from \emph{top left} to \emph{bottom right}, respectively). The comparison data set includes the works of \citet{Andrews17} at all bands, \citet{Dahlen07} and \citet{Tresse07} at the FUV, NUV and $B$-bands, and finally \citet{Budavari05} for the FUV and NUV. Additional data sets in the FUV are \citet{Bouwens16}, \citet{Cucciati12}, \citet{Reddy09} and \citet{Schiminovic05}. For the $B$-band we add the \citet{Marchesini07} and \citet{Gabasch04} data. And finally, we integrate the $K$-band LFs of \citet{C10} and \citet{Mortlock16}, and include the \citet{Pozzetti03} measurements.}
\label{fig:ld}
\end{figure*}

\begin{figure*}
    \includegraphics[width=0.475\textwidth]{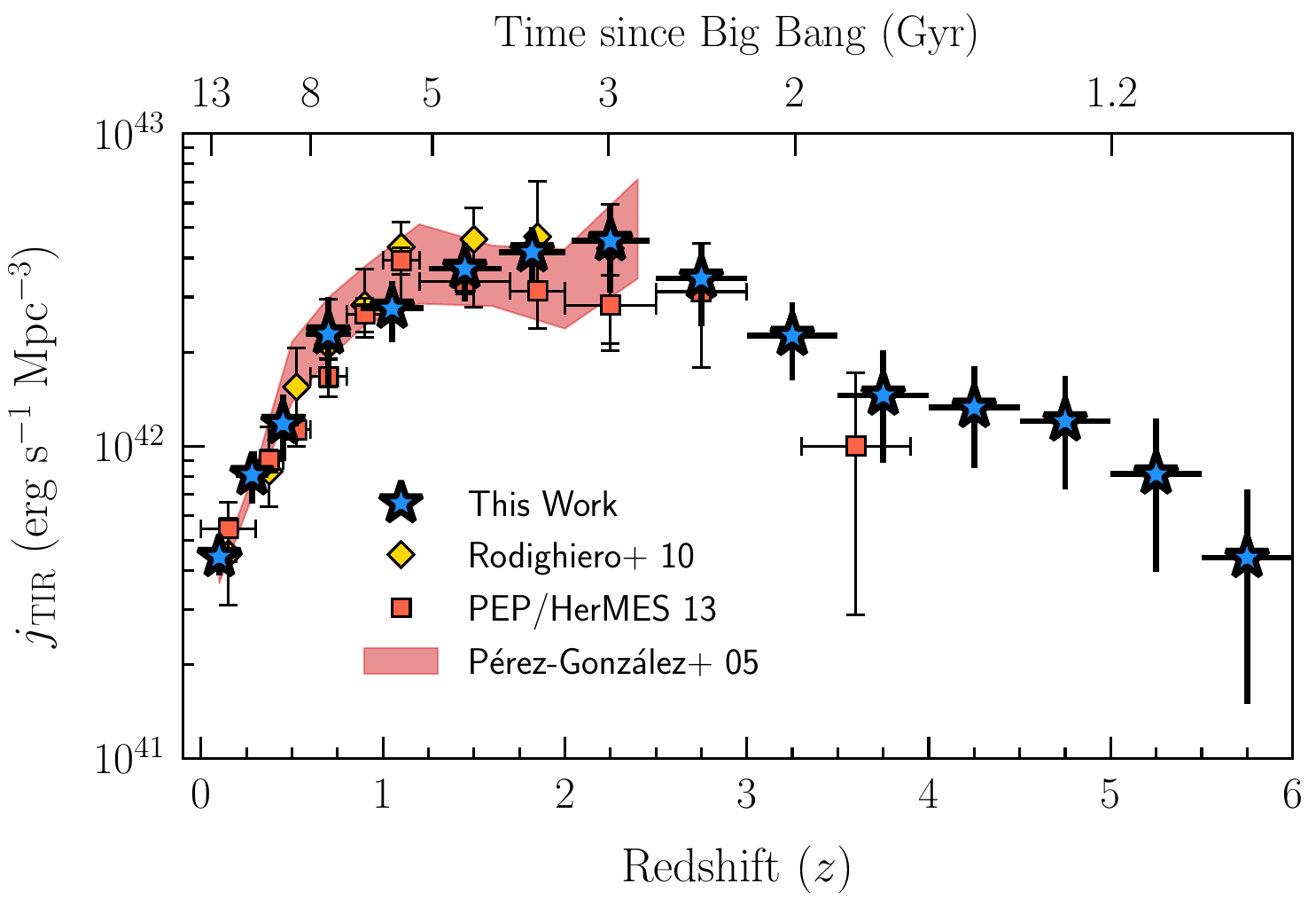}
    \hspace{0.25cm}
    \includegraphics[width=0.475\textwidth]{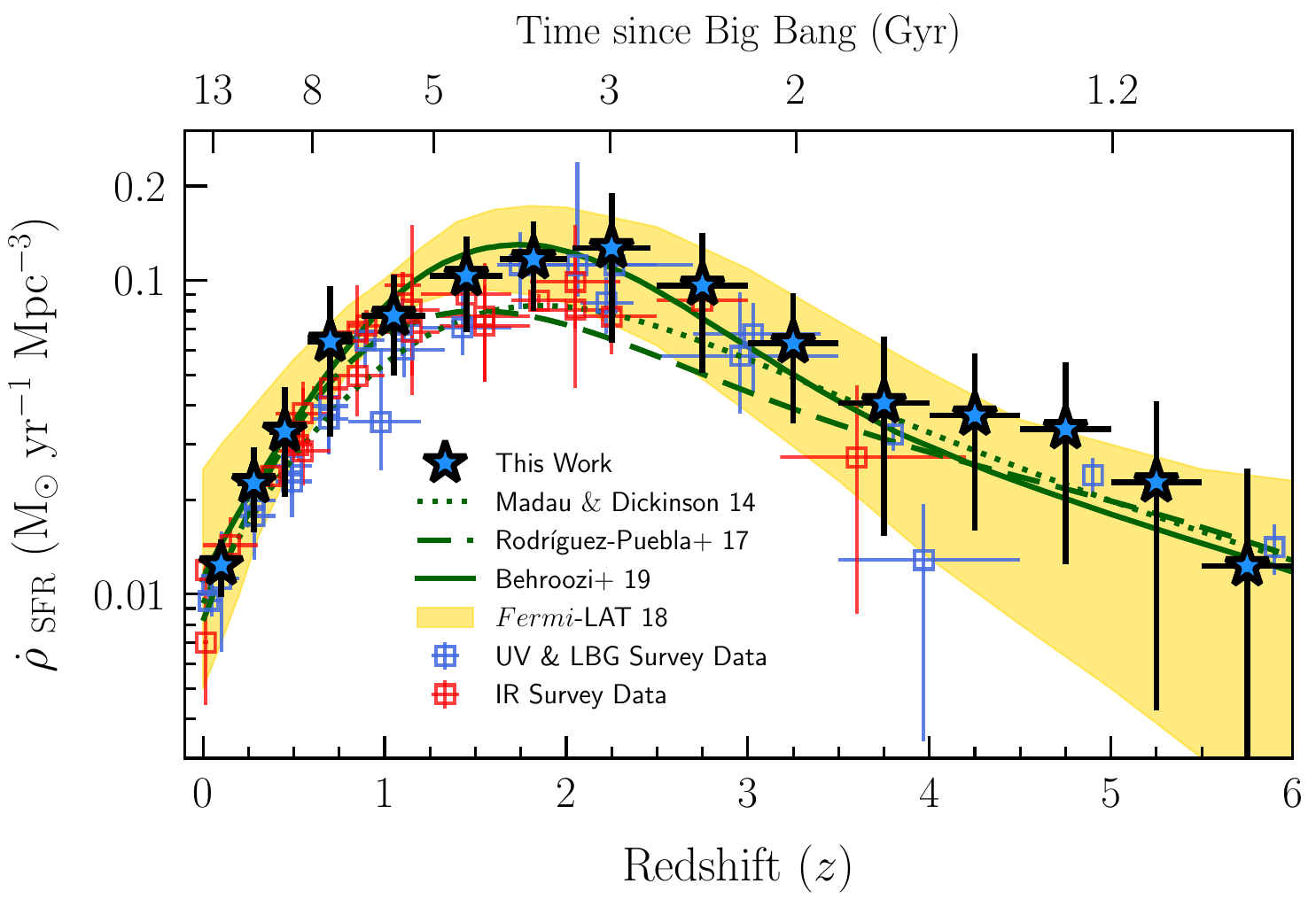}
\caption{Cosmic star-formation history (CSFH) derivation. \emph{Left}: our total-IR (TIR) luminosity density at 8-1000$~\mu$m. The values from \citet[PEP/HerMES 2013][]{PEP13}, \citet{Rodighiero10} and \citet{PG05} are also plotted for comparison. \emph{Right}: CSFH derived using our UV and TIR luminosity density estimations. The UV+IR data compilation of \citet{MD14} are given together with the CSFH parametrization of the same paper (red line). Finally, the gamma-ray attenuation determination of the CSFH derived in \citet{Fermi18} is shown as our fiducial comparison study (green shaded area). All values have been converted to a Chabrier IMF.}
\label{fig:ld_tir_csfh}
\end{figure*}

\section{Methodology}\label{sec:method}
\subsection{Galaxy emissions in the mid- and far-infrared}\label{sub:method_ltir}

In this paper, we aim at estimating the EBL at all wavelengths from the UV to the FIR as a function of redshift up to $z\sim6$. Concerning the IR calculations, we have to take into account that a fraction of galaxies in a stellar mass limited sample such as ours are not actually detected by the deepest surveys carried out with {\it Spitzer} and Herschel telescopes. In order to cope with this obstacle, we have used an elaborate method to assign a dust emission SED in the mid- and far-infrared to every single galaxy. This method is based on actual flux measurements when available and on attenuation estimations and an energy-balance argument for sources below the detection limit of the IR surveys. We describe the method in detail in the following paragraphs.

In our EBL sample, 4,071 galaxies have simultaneous detections in {\it Spitzer}/MIPS+Herschel bands, \ie $\sim 2.6$\% out of the total number of 154,447 galaxies, according to the prior-based catalogs in \citetalias{Barro19}. For these sources, the IR SED can be recovered with high confidence. For another 5,391 galaxies (3.4\% of the entire sample), the reddest detected photometric point is MIPS/24~$\mu$m. There is no FIR detection for the rest of 144,985 ($\sim 93.8\%$). Given their large numbers and an imhomogeneous redshift distribution of the IR detections, their contribution to the IR background might be significant. In order to account for the whole EBL sample in our analysis, we consider 3 types of sources: (1) MIPS+Herschel detections, (2) MIPS/24~$\mu$m-only detections, and (3) no-FIR detections.

For type (1) where there is detection by both MIPS and at least one Herschel band, we take the best-fit \citetalias{ChEl01} template given by Rainbow to reproduce the IR-SED. Having a mid-to-far IR color allows us to obtain robust flux interpolations in the spectral region where the dust emission peaks and highly accurate estimations of the total infrared luminosity (L(TIR) hereafter, see \autoref{fig:seds}). L(TIR) is defined as the integrated luminosity between 8 and 1000~$\mu$m.

In case (2), sources where the reddest detection is MIPS/24~$\mu$m, the global dust emission has to be extrapolated from a single point. This extrapolation must be carried out with care since galaxies at different redshifts might present different warm-to-cold dust characteristics. Indeed, it is well known that Ultra-Luminous Infrared Galaxies (ULIRGs) at $z\sim2$ present more prominent PAH features (probed by the MIPS datapoint) than galaxies of the same luminosity in the local Universe. High-$z$ ULIRGs present SEDs which are more similar to local LIRGs, with strong PAH emission \citep[see][]{Papovich11}. We note, however, that we have far-IR detections for ULIRGs at high-redshift, so this effect is included in our data. For less luminous objects, systematic uncertainties are lower \citep[see][ \citetalias{Rieke09}]{Sajina12}. In any case, we obtain a dust emission model for every MIPS-only detection by first estimating a L(TIR) with methods that account for the redshift evolution of (U)LIRGs, and then assigning a model that mataches that luminosity. For the first part of our method, we use the expression from \citet[hereafter \citetalias{Wuyts11}]{Wuyts11} that derives the L(TIR) given the observed MIPS/24~$\mu$m flux density in mJy ($\mathrm{F_{\nu}}$), using a coefficient C($z$) that depends on the redshift of thegalaxy, 
\begin{equation}\label{eq:W11}
    \mathrm{L(TIR)~[L_{\odot}] = C(z) \times F_{\nu}[mJy]},
\end{equation}

\noindent
where L$_{\odot}$ corresponds to solar bolometric luminosity, for which we adopt L$_{\odot}$ = 3.826 $\times 10^{33}$ erg s$^{-1}$. Then, we associate a scaled \citetalias{ChEl01} dust emission template to every inferred L(TIR). We choose the \citetalias{ChEl01} templates because they are normalized by L(TIR) luminosity, so that there is a direct association between L(TIR) and a \citetalias{ChEl01} template. We will discuss the implications of this choice in Sect.\ \ref{sec:results}. \autoref{fig:seds} shows an example of a galaxy of type (2), depicting in red the final dust-emission model. We note that our method also includes checking that the model is consistent with the upper limits imposed by the Herschel data, typically sampling the peak of the IR SED.

For type (3) sources, those with no detection in the IR, we perform an analysis of the attenuation by using the total SFRs presented in \citetalias{Barro19}. In this paper, a calibration of the attenuation versus the UV-slope is carried out using the IRX-$\beta$ relationship \citep{Meurer99} as a function of redshift and taking into account upper limits imposed by {\it Spitzer} and Herschel. The IRX-$\beta$ relationship is found to be dependent of the total SFR, which is correlated with stellar mass via the main-sequence plot. In summary, different curves are used for different galaxies in order to obtain robust estimations of the attenuation even for galaxies below the MIPS and Herschel detection limits. Using these curves, an attenuation is assigned to every galaxy in the CANDELS surveys (and in our EBL sample, selected from them) and total SFRs are estimated starting from UV-based fluxes and slopes. For further discussion, we refer the reader to \citet{DSanchez16} and \citet{RMunoz19}.

Total SFRs for every CANDELS galaxy are calculated in \citetalias{Barro19} by adding the contribution from TIR and the observed UV emission (see Eq.\ \ref{eq:sfr_uvtir}). This approach uses the Kennicutt relations \citep{Kennicutt1998} and assumes a Chabrier IMF \citep{Chabrier2003}:

\begin{equation}\label{eq:sfr_uvtir}
\begin{split}
    \mathrm{SFR} & = \mathrm{SFR(UV) + SFR(TIR)},\\
    \mathrm{SFR[M_{\odot} yr^{-1}]} &= 1.07 \times 10^{-10} \left(3.3 \times \mathrm{L(2800) + L(TIR)} \right) / \mathrm{L_{\odot}},
\end{split}
\end{equation}

\noindent
where the observed rest-frame monochromatic luminosity at 2800\AA~wavelength, L(2800), is chosen as representative of the unobscured UV SFR (uncorrected for dust extinction). L(TIR) is calculated integrating the galaxy--SED over the 8 to 1000~$\mu$m wavelength range.

For galaxies not detected in the IR, \citetalias{Barro19} provides a total SFR estimated from UV fluxes with the method described above, and an `observed' SFR, obtained directly from the UV emission. Using these 2 quantities we can obtain the amount of star formation hidden by dust, which can then be translated to a TIR luminosity, and that can be used to assign a template to each galaxy. In this procedure, we again take into account the 5$\sigma$ upper limits in all the FIR bands (\ie MIPS/24 and 70~$\mu$m and Herschel PACS/100 and 160~$\mu$m, SPIRE/250, 350 and 500~$\mu$m, see \autoref{tab:IRuppLIM}), so our final template is consistent with observations. 

For galaxies in case (3), we then first compute a SFR(UV) excess at 2800\AA~($\Delta \mathrm{SFR(2800)}$),
taking into account the difference between corrected and observed SFR(2800) values, \ie
\begin{equation}\label{eq:sfr_excess}
   \Delta \mathrm{SFR(2800) = SFR(2800)_{corr} - SFR(2800)_{obs}}.
\end{equation}
This excess indicates the fraction of UV radiation absorbed by dust. Assuming an energy-balance, that hidden star formation will be revealed in the IR, where we should be able to recover the SFR excess if we had deep enough data. Using the Kennicutt relation and assuming a Chabrier IMF, we can calculate L(TIR) from $\Delta \mathrm{SFR(2800)}$ as:
\begin{equation}\label{eq:ltir_delta}
    \mathrm{L(TIR) (}\mathrm{erg~ s}^{-1}\mathrm{)} = 3.8 \times 10^{43} \Delta \mathrm{SFR(2800)}.
\end{equation}

Finally, we can associate a scaled \citetalias{ChEl01} dust emission template to every L(TIR), similar to the procedure for galaxies in case (2). In \autoref{fig:seds}, we also show an example of this method. Note that the IR SED is roughly compatible with the 30~$\mu$Jy MIPS/24~$\mu$m upper-limit flux density. 

\autoref{fig:ltir_hist} shows the L(TIR) distribution for the entire EBL sample before (\ie assuming 5$\sigma$ IR upper limits when appropriate) and after our own L(TIR) estimation for the three galaxy types. In addition, the same distribution is shown for the (1) and (2) cases, separately in the inset. Finally, the L(TIR) evolution with redshift is plotted in \autoref{fig:ltir_z}, where (1), (2), and (3) are again highlighted. There is a cluster of galaxies placed in between $z=0$ and $z=4~$ ($\sim 5\%$ out of the total number of sources over the same redshift range) whose L(TIR) values are in between 10$^{5-6}$~L$_{\odot}$. We studied and quantify the emission of these outliers and found that their contribution to the global IR background is within the uncertainties of our EBL model. 

\subsection{Extragalactic background light model formalism}\label{sub:method_model}
The evolving EBL-SED can be obtained from stacking individual galaxy-SEDs and calculating the total luminosity density at different redshift bins. 

The comoving luminosity density  $j(\lambda, z)$ can be interpreted as the energetic output of the Universe in a given comoving volume. It results from the emission of all the galaxies enclosed within the same volume, and it is measured in terms of total luminous power per unit comoving volume and frequency interval (erg s$^{-1}$ Mpc$^{-3}$ Hz$^{-1}$). The monochromatic (\ie at a given photometric band~/~central wavelength) luminosity densities at a given redshift range can be calculated by integrating the corresponding luminosity function (LF) at the same band and redshift range. Equivalently, we can estimate the luminosity density SED as a function of wavelength by adding all the galaxy-SEDs in the same redshift interval. The last approach is what we use in this work.

We divide our EBL galaxy sample into the following 15 redshift bins: $\Delta z_i=z_{i+1}-z_{i}$, where $z_i$ = (0, 0.2, 0.35, 0.55, 0.85, 1.25, 1.65, 2, 2.5, 3, 3.5, 4, 4.5, 5, 5.5, 6).  The bins were chosen to have approximately the same cosmological time interval on each bin.  We then stack together all the rest-frame galaxy-SEDs ($L_{\nu}(\lambda,z_i)_j$) contained in the same redshift bin ($\Delta z_i$). Finally we divide by the comoving cosmological volume ($V_c(z_i)$), and scale by the total solid angle ($\mathrm{A_{\Omega}}$) covered by the CANDELS survey over the whole sky:
\begin{equation}\label{eq:ld}
    j(\lambda, \Delta z_i) = \dfrac{\sum_{j = 1}^{n} L_{\nu}(\lambda,z_i)_j} {V_c(\Delta z_i) \cdot \mathrm{A_{\Omega}}},
\end{equation}

\noindent
where $n$ represents the number of galaxies at the $\Delta z_i$ redshift interval. The comoving volume is defined, assuming a $\Lambda$CDM cosmology, as:
\begin{equation}\label{eq:com_vol}
    V_c(z_i) = \dfrac{4 \pi c^3}{3 H_0^3} \left(\int_{0}^{z_i} \dfrac{dz'}{\sqrt{\Omega_\Lambda + \Omega_M (1+z')^3}}\right)^3,
\end{equation}

\noindent
with $c = 2.99 \times 10^8$~m/s, the speed of light in vacuum, and $V_c(\Delta z_i) = V_c(z_{i+1}) - V_c(z_i)$. Finally, the EBL intensity at a certain wavelength $\lambda$ and redshift $z_i$ is produced by the contribution of light at shorter rest-frame wavelengths emitted in the \emph{light path} since $z_{max}=6$ to the current $z_i$. That is, the farther away we look, the shorter the rest-frame wavelength that contributes to our comoving EBL-SED, specifically by a $(1+z)/(1+z')$ contracting factor: $z < z' < z_{max}$. Formally, this EBL-SED can be recovered from the evolving comoving luminosity density integration as follows \citep{MoVdBoschWhite2010},
\begin{equation}\label{eq:EBL}
    \lambda I_{\lambda}(\lambda, z_i) = \dfrac{c^2}{4\pi\lambda} \int_{z_i}^{z_{max}} j\left(\lambda(1+z_i)/(1+z'), z')\right) \left|\dfrac{dt}{dz'}\right| dz',
\end{equation}

\noindent
where $\lambda I_{\lambda}(\lambda, z_i)$ is given in nW m$^{-2}$ sr$^{-1}$ units and the corresponding $j(\lambda, z_i)$ function can be interpolated over an appropriate $\delta{z}$ step size vector, if needed. The $|dt/dz'|$ factor comes from the adopted $\Lambda$CDM cosmology:
\begin{equation}\label{eq:cdm_factor}
    \left|\dfrac{dt}{dz'}\right| = \dfrac{1}{H_0(1+z')\sqrt{\Omega_{\Lambda} + \Omega_M (1+z')^3}}.
\end{equation}

Uncertainties in the EBL-SED determination are calculated from every single galaxy SED, by varying the flux density of every photometric point over its 1$\sigma$ error, and allowing the associated templates to scale with the errors in a Monte-Carlo approach (using 100 iterations for computational reasons). At the end, we have a 1$\sigma$ band uncertainty for every galaxy-SED between 0.1 and 1000~$\mu$m. For sources with only MIPS/24~$\mu$m detection at the IR, we re-scale the IR templates only within the 24~$\mu$m flux density error bars, whereas we do not assume any error at the IR bands for the cases of non-detected IR sources (\ie sources whose IR-SED has been obtain from the SFR(UV)-excess). 

These SED uncertainties are properly propagated to the luminosity density through Eq.\ \ref{eq:ld}, and then to the EBL-SED using Eq.\ \ref{eq:EBL}.

\subsection{Completeness corrections}\label{sub:method_completeness}

As we move to higher redshifts we probe less and less of the low-mass end of the stellar mass function (SMF), and the contribution from galaxies below the detection limit can be significant at the highest redshifts considered here ($z>3$). 


In order to assess a completeness correction, we use two SMFs from the literature \citep{Ilbert13,Grazian15}, and assumed SEDs for the galaxies below the detection limit similar to the sources detected just above it. We explain the method in detail in the following paragraphs.


First, and for every redshift bin $\Delta z_i$, we compute the expected number of sources below the mass completeness limit $\phi_{M_*}(M_* \leq M_*^{lim}, z_i)$, using the \citet{Ilbert13} ($0 \leq z < 4$) and \citet{Grazian15} ($4 \leq z < 6$) SMFs, as appropriate depending on the considered redshift range, and down to a given mass threshold of $M_*=10^5 ~M_{\odot}$. An example of the mass-density of object estimated by the \citet{Ilbert13} SMF is shown in \autoref{fig:smf} for the $0.35 \leq z < 0.55$ redshift bin. We also perform and use our own single-Schechter fits to the SMF at those redshifts where the above mentioned SMF don't match the data really well (more importantly at the lowest redshifts).

Secondly, we build UVJ color-color diagrams for every redshift bin \citep{Williams09, Whitaker11}, and compute the fraction of quiescent ($f_Q$) and star-forming galaxies ($f_{SF}$) within the 0.5~dex mass bin immediately above of the corresponding completeness limit. We also assume that the empirical UVJ relation that separates quiescent from SF galaxies at the lower redshifts is still valid at $z>3$.

Then, we compute the average SED ($\langle L_{\nu}(\lambda,z_i) \rangle _{M_*^{lim}}$) for these galaxies at the mass limit as the weighted median SED of quiescent and SF galaxies analyzed before; that is: \begin{equation}\label{eq:mean_sed}
\begin{split}
    \langle L_{\nu}(\lambda,z_i) \rangle_{M_*^{lim}} =& f_{Q,i}\times \langle L_{\nu}(\lambda,z_i)_Q \rangle + \\ &+f_{SF,i}\times \langle L_{\nu}(\lambda,z_i)_{SF} \rangle.
\end{split}
\end{equation}

Following this approach, the additional contribution to the total luminosity density of galaxies below the mass completeness limit is:
\begin{equation}\label{eq:jMlim}
\begin{split}
    j(\lambda,z_i)_{\leq M_*^{lim}} = & ~\langle L_{\nu}(\lambda,z_i)\rangle _{M_*^{lim}, z_i} \\
    & \times \int_{10^7~M_{\odot}}^{M_*^{lim}} \phi_{M_* \leq M_*^{lim}, z_i} ~dM_*, \\
\end{split}
\end{equation}

\noindent 
where the integral comes from the SMF definition: $d \phi_{M_*} = dN / (dV_c d M_*)$, with $N$ being the number of galaxies per mass bin and comoving volume element. Therefore, the total corrected luminosity density ($j(\lambda,z_i)$) is the sum of both the observed ($j(\lambda,z_i)_{obs}$, obtained by the stacking of galaxies up to every limiting mass) and this additional contribution of low-mass galaxies; \ie:
\begin{equation}\label{eq:jtot}
    j(\lambda,z_i) = j(\lambda,z_i)_{obs} + j(\lambda,z_i)_{\leq M_*^{lim}}.
\end{equation}

This is, indeed, the luminosity density that has to be introduced in Eq.\ \ref{eq:ld} and \ref{eq:EBL}. However, the effect of our completeness corrections on the EBL SED is noticeable only at the higher redshifts. At $z=0.1$ the contribution from the corrections to the total integrated EBL is only of 6\%, at $z=1$, $z=3$, $z=5$ is 10\%, 15\%, and 20\%, respectively. The uncertainties on these completeness corrections are included in the luminosity densities by adding in quadrature the difference between the corrected and non-corrected (observed) luminosity densities at every wavelength.


%% file: 4_results.tex
\begin{figure*}
    \includegraphics[width=0.85\textwidth]{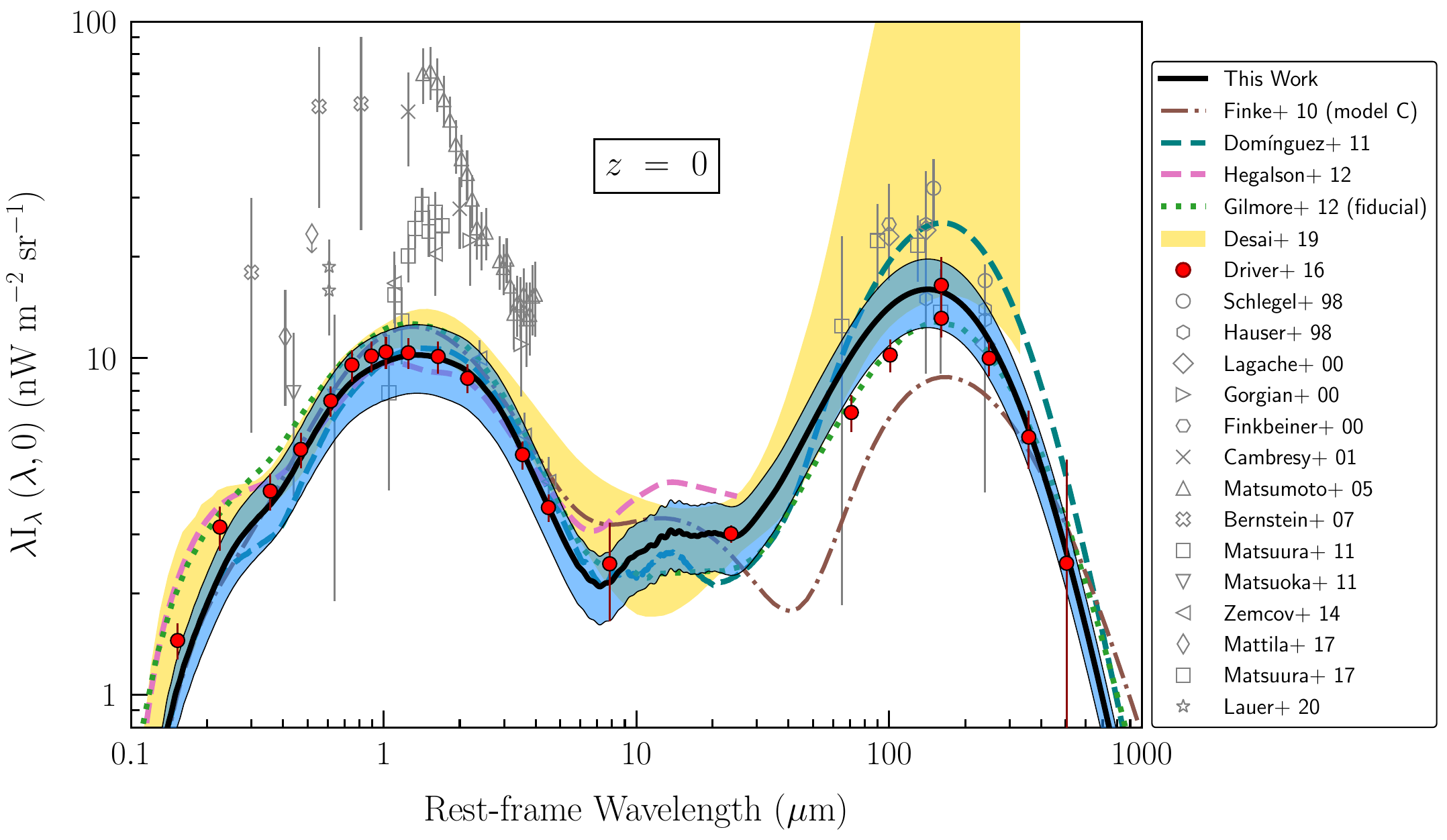}
\caption{The local ($z=0$) spectral energy distribution of the EBL (black solid line, 1$\sigma$ model uncertainties is enclosed within the blue shaded area). The green dotted and brown dashed-dotted lines are the semianalytical and phenomenological models of \citet{Gilmore12} and \citet{F10}, while the dashed blue and pink lines corresponds to the \citet{D11} and \citet{H12} empirical curves, respectively. The yellow band spans over the 1$\sigma$ limits for the local EBL determination of \citet{Desai19}, from blazars' gamma-ray attenuation spectra. Finally, the red filled points come from galaxy number counts in \citet{D16}. Open symbols constitute a compilation of diverse direct measurements from the literature (see text).}
\label{fig:ebl_local}
\end{figure*}

\begin{figure*}
    \includegraphics[width=0.90\textwidth]{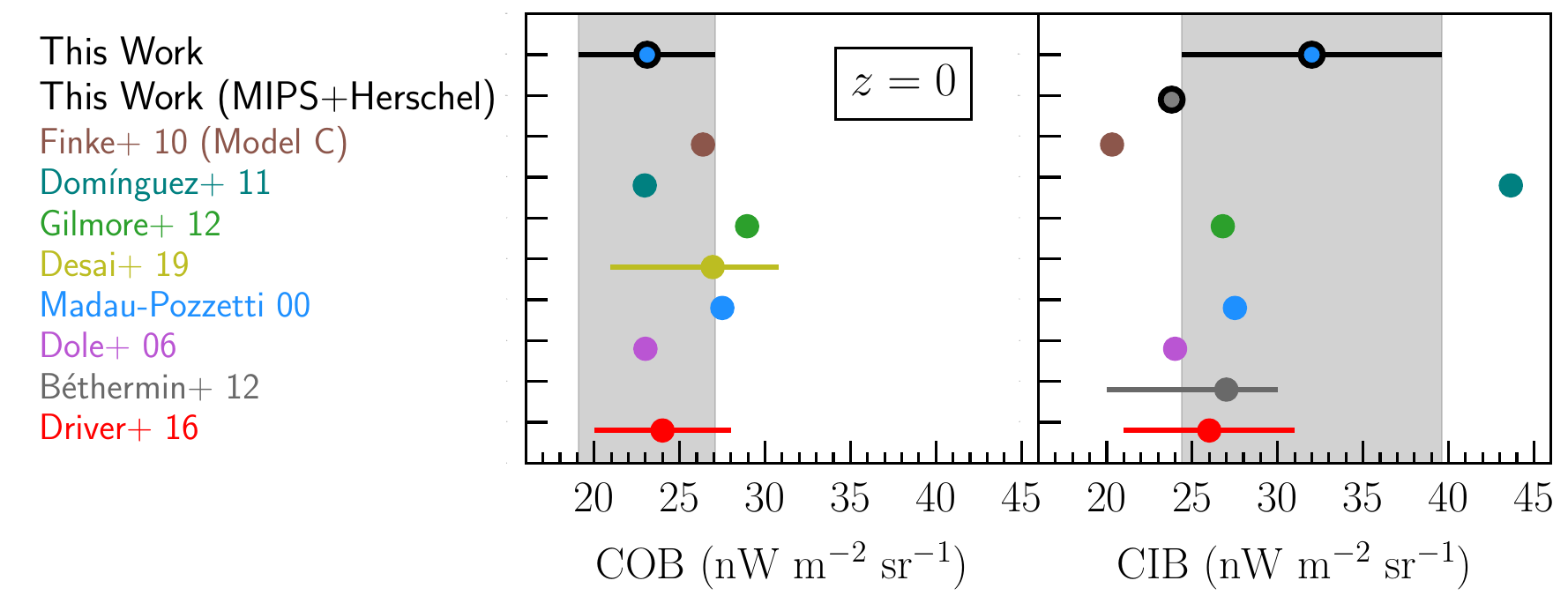}
\caption{Comparison between different EBL models and measurements of the Cosmic Optical (COB, 0.1-8$~\mu$m) and Infrared Backgrounds (CIB, 8-1000$~\mu$m) at $z=0$ \citep[similar to][]{D16}. The results from This Work are shown in black, while in brown, blue, green and yellow are the results from \citet{F10}, \citet{D11}, \citet{Gilmore12} and \citet{D19} models, respectively. The results from \citet{MadauPozzetti00}, \citet{Dole06}, \citet{Bethermin11} and \citet{D16} galaxy number counts are chronologically  plotted in light-blue, purple, grey and red. We also show the measured CIB if only combined MIPS+Herschel detections are included in our work, as discussed in the text. Uncertainties have been included only when available in the literature.}
\label{fig:ebl_cob_cib}
\end{figure*}

\begin{figure*}
    \includegraphics[width=0.80\textwidth]{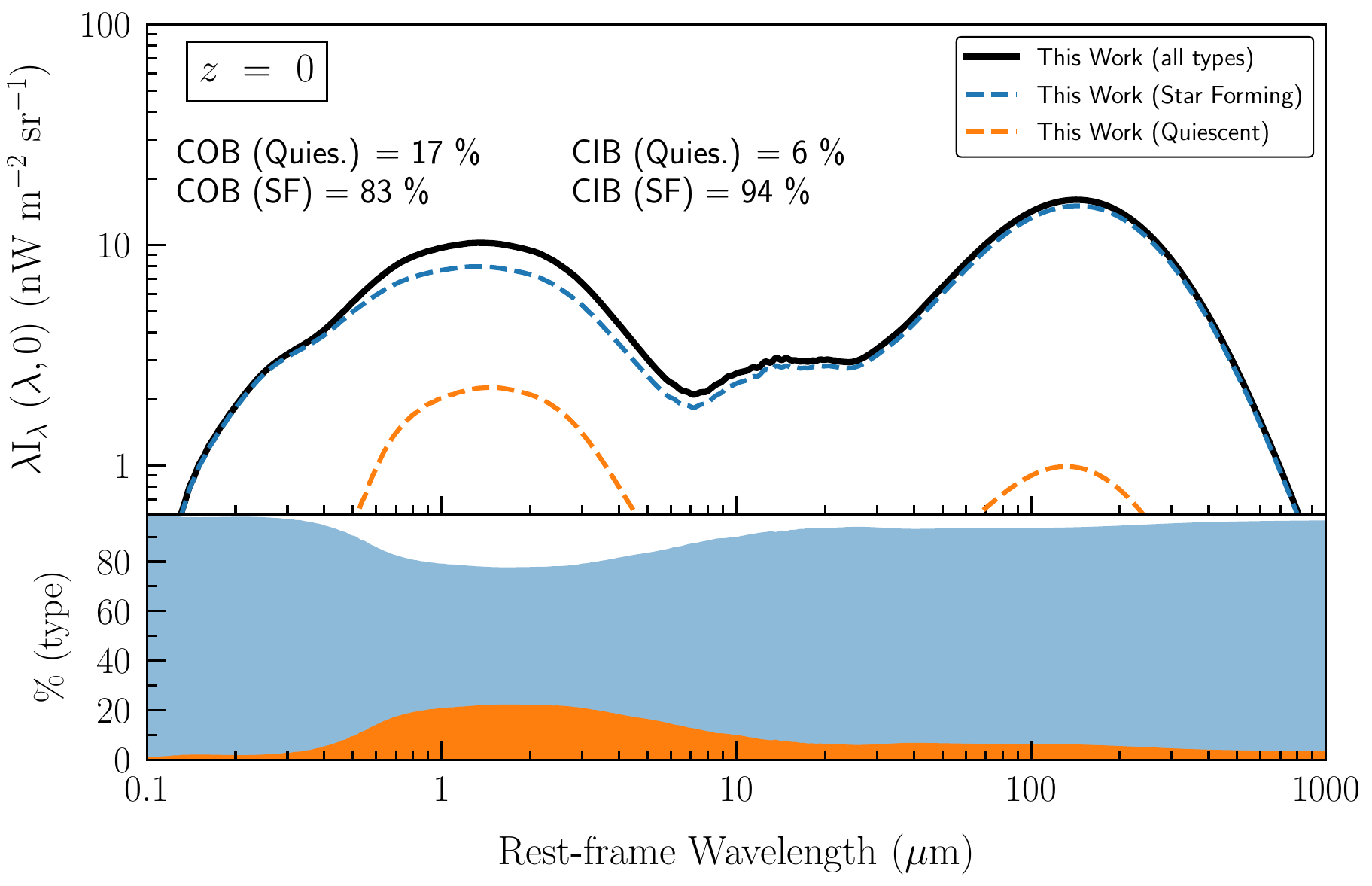}
\caption{Relative contribution to the local EBL intensity of quiescent (orange) and star-forming galaxies (blue), under UVJ selection criteria. Star-forming contribution dominates both the COB and the CIB with a $\sim 70\%$ and $\sim 90\%$ of the integrated EBL intensity, and their contribution increases rapidly with redshift. The quiescent contribution is mainly due to passive galaxies at $z<1$.}
\label{fig:ebl_mass_qsf}
\end{figure*}

\begin{figure*}
    \includegraphics[width=0.99\textwidth]{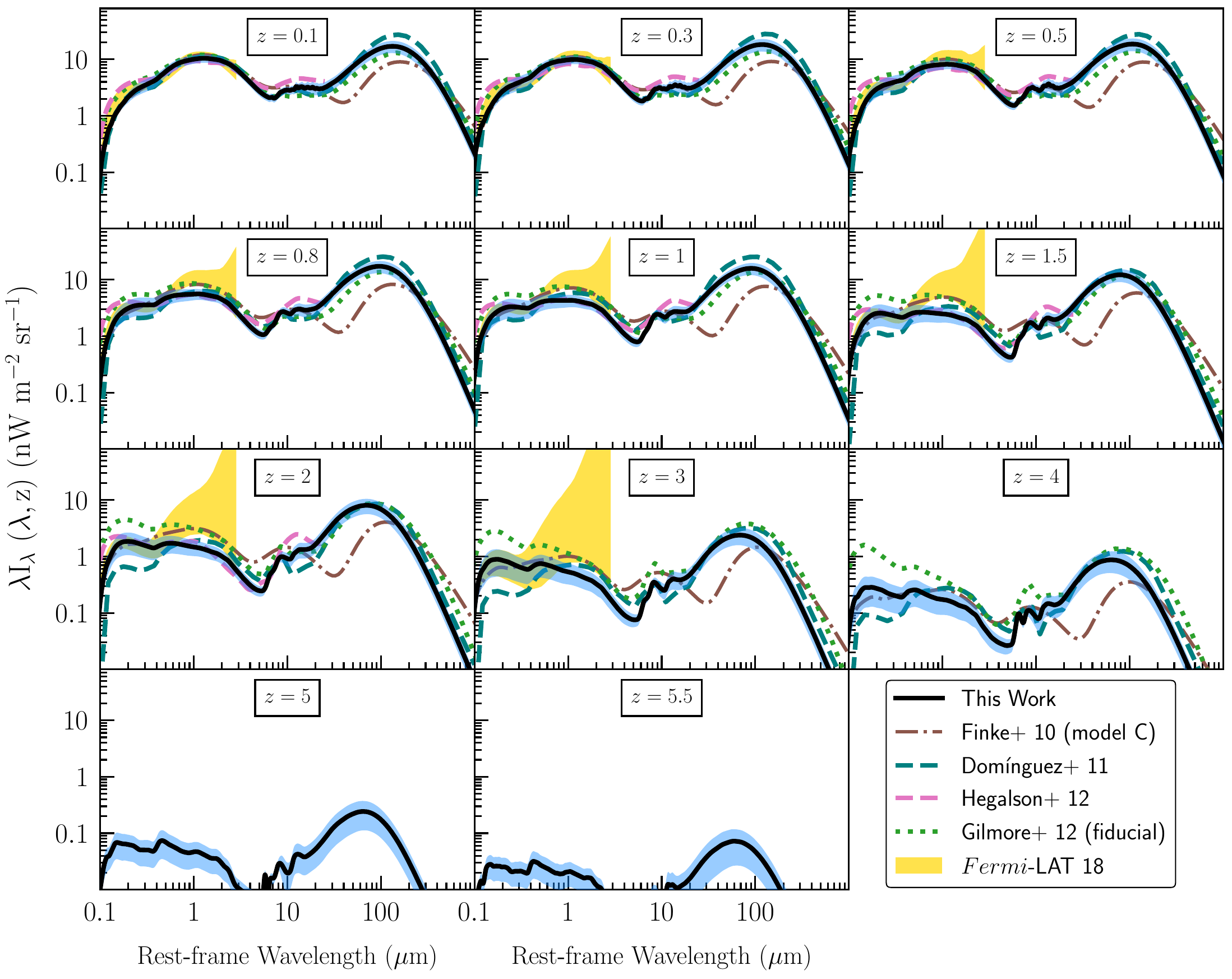}
\caption{Evolving SED of the EBL between $z=0$ and $z=6$ in the different redshift snapshots of our model. Also plotted are the results of the \citet{F10}, \citet{D11}, \citet{H12} and \citet{Gilmore12} models, with the same styles as in \autoref{fig:ebl_local}. The yellow band correspond to the results of \citet{Fermi18}, interpolated between $z=0$ and $z=3$. Each model has been extended up to its particular maximum redshift, and interpolated to the intermediate redshift values.}
\label{fig:ebl_z}
\end{figure*}

\section{Results and discussion}\label{sec:results}
\subsection{Luminosity densities and cosmic star formation history }\label{sub:result_LDs}
Following the formalism described in the previous section, we derive the monochromatic luminosity density in different bands, and compare our results with a set of measurements from the literature. In particular, we compare with FUV (1500~\AA), NUV (2800~\AA), optical ($B$-band, 4400~\AA) and NIR ($K$-band, 2.2$~\mu$m) luminosity densities between $z=0$ and $z=6$ (see \autoref{fig:ld})\footnote{Results from this work can be found at \url{https:www.ucm.es/blazars/ebl}.}

The FUV (1500~\AA) and NUV (2800~\AA) luminosity density results are shown together with the values of \citet{Budavari05} and \citet{Andrews17} at low-$z$; and \citet{Reddy09} and \citet{Bouwens16} at high-$z$. Spanning the entire redshift range we compare with \citet{Cucciati12}, \citet{Dahlen07}, \citet{Tresse07} and \citet{Schiminovic05}. Our calculations and the literature agree within 1$\sigma$ at all redshifts, excepting the FUV estimations from Lyman-Break Galaxies surveys \citep{Reddy09} at $z \sim 3-4$, where our points fall systematically by $\times$3 factor above.

In \autoref{fig:ld}, the $B$-band luminosity density (4400~\AA) is compared with \citet{Tresse07} and \citet{Dahlen07}, and \citet{Andrews17}. Additional data compilations at high-$z$ come from \citet{Marchesini07} and \citet{Gabasch04} works. Again, our concordance with the literature up to $z=1$ is noticeable, whereas at $1 \le z <2$, our data fall systematically above most of the optical-LF integrations by a factor of $\times 1.5$. Beyond $z \ge2$, comparison fails because of the different values between the different measurements reported in the literature. Our estimates at $z>2$ are within 1$\sigma$ agreement with \citet{Gabasch04} and a factor $\times 3$ above \citet{Marchesini07} at $z \sim 2-3$.

The $K$-band luminosity density is presented in \autoref{fig:ld} together with the \citet{Andrews17} and \citet{Pozzetti03} published values at low-$z$. We also include the results from the integration of the \citet{Mortlock16} and \citet{C10} $K$-band LFs. We are compatible within uncertainties with work of \citet{Mortlock16} and \citet{Andrews17}. Our NIR intensities are systematically above the values given by \citet{Pozzetti03} and \citet{C10} at all redshifts. The reason is that these studies are shallower in fluxes than ours.


An observable directly related with the CSFR density is the TIR luminosity density. Our $j_{TIR}$ is shown in \autoref{fig:ld_tir_csfh} over $0 \leq z < 6$, as well as the estimations from \citet{PG05}, \citet{Rodighiero10}, and \citet[PEP/HerMES 2013]{PEP13}. Note that the agreement between the different measurements is good at all redshifts.

Finally, the cosmic star-formation history (CSFH) can be computed following Eq.\ \ref{eq:sfr_uvtir}, using the NUV and TIR luminosities. Our CSFH between $0 \leq z < 6$ is shown in \autoref{fig:ld_tir_csfh} in comparison with the \citet{MD14}, \citet{aldo17}, \citet{Fermi18} and \citet{behroozi19} results (all results have been converted to a Chabrier IMF when needed). We stress that our result is in agreement within errors up to $z\sim 6$ with those from $\gamma$-ray attenuation. Since the $\gamma$-ray technique is sensitive to all light produced by galaxies, even those too faint that escape the detection in galaxy surveys, we conclude that our model, including completeness correction, is correctly reproducing most of the galaxy contribution.


\subsection{Spectral energy distribution of the EBL}

\subsubsection{The local extragalactic background light}\label{sub:result_EBLlocal}
Applying Eq.\ \ref{eq:EBL} to our total luminosity density, we can estimate the evolving EBL. To do that we have interpolated the set of 15 luminosity density SED into a 2-dimensional ($z, \lambda$) grid with $\delta z = 0.1$ steps in the $0 \leq z < 6$ interval, and a common spectral synthetic resolution of $\delta (\log \lambda [\mu m]) = 0.01$ in the 0.1--1000~$\mu$m range. 

\autoref{fig:ebl_local} shows the local EBL SED. This SED is compared with the results from other methodologies such as direct detection \citep{Schlegel98, Hauser98, Lagache00, Gorjian00, Finkbeiner99, Cambresy01, Matsumoto05, Bernstein07, Matsuura11, Matsuoka11, Zemkov14, Mattila17, Matsuura17, Lauer20}\footnote{A re-analysis of the Zodiacal light beyond the Earth orbit detected by {\it Pioneer 10} \citep{Matsumoto18}, suggests that instrumental offsets affected the \citet{Matsuura11} measurements.}, galaxy counts \citet{D16}, $\gamma$-ray attenuation \citep{Desai19} and other EBL models \citep{F10,D11,H12,Gilmore12}. 

\autoref{fig:ebl_local} shows that our model is $1\sigma$ compatible with the galaxy count data from \citet{D16}. The largest differences between the results from several models occur at the IR. \citet{D11} estimate a larger IR background by factor of $\times$2 at the IR peak. This was a possibility discussed in the paper since they did not have FIR photometry and that spectral region came from extrapolations. On the other side, \citet{F10} estimate a lower FIR contribution at $\sim 200~\mu$m by a $\times$1.5 factor. 

In the MIR coverage, although our curve is compatible with \citet{Bethermin11} at 24~$\mu$m, and marginally compatible with \citet{Dole06} within 1$\sigma$, our result sits above most of the models in the 10-20~$\mu$m range (excepting \citet{H12}). This may be because of the chosen \citetalias{ChEl01} IR templates. Since they are calibrated with low-$z$ galaxy observations, they included a higher level of PAHs emission, which could raise up the EBL intensity at those wavelengths in our calculation.

Most of the direct detection data are not compatible with our estimates. It is possible that these data, at least, in the optical and IR, can be strongly affected by zodiacal light contamination. However, there are some exceptions. 

The more sophisticated approach by \citet[][the ``Dark Cloud shadow'' method]{Mattila17} is higher but compatible within $2\sigma$ with our calculations at 0.4~$\mu m$ ($11.6 \pm 4.4$~nW m$^{-2}$~sr$^{-1}$, see previous reference for more details). \citet{Zemkov14} used fluctuations in the pixel intensity power spectrum to measure the EBL at 1.1 and 1.6~$\mu m$. They report a slightly higher (but still compatible at 1.1~$\mu m$) EBL intensity with respect to us: $16.7^{+5}_{-4}$, $20.4^{+6}_{-5}$ ~nW m$^{-2}$~sr$^{-1}$, respectively \citep[see also][]{Matsumoto20}. These techniques are less biased by zodiacal foreground contamination. Finally, the recent work by \citet{Lauer20} using {\it New Horizons} measured the EBL at 0.6~$\mu m$ using direct detection at a distance of around 40~AU from the Earth. At that distance, zodiacal light is much fainter than closer to the Sun. Therefore, \citet{Lauer20} constitutes probably one of the most accurate single band local EBL measurements up to date ($15.9 - 18.7$~nW m$^{-2}$~sr$^{-1}$). 

Although our results are mainly consistent within $2\sigma$ with the former studies, the higher values reported in the literature may indicate that either (1) existing galaxy catalogs are missing very low surface brightness sources and/or (2) completeness corrections are not accounting for the entire contribution of the SMF faint-end galaxy population (e.g. \citet{Conselice16} reported $\sim 47.2$~nW~m$^{-2}$~sr$^{-1}$ at $\sim 0.8~\mu m$ from LF extrapolations, although this high value may come from overestimations in the number of such faint sources).

Relative to $\gamma$-ray attenuation based studies, they may not be sensitive to the $z \leq 0.1$ fainter galaxies since the optical depth to such sources is very low. A comparison between the results from direct detection (absolute photometry methods) and for instance \citet{Desai19} may indicate that $\gamma$-ray observations could be missing a light component that is being detected by the more direct approaches.

Finally, we estimate the influence of our sample selection in the local EBL intensity. As discussed in Sect.\ \ref{sub:data_filtering}, the strongest cut in our sample is $S/N=5$ at the HST/WFC3/F160W band. We calculate F160W galaxy number counts imposing $S/N=2$ (see \autoref{fig:ncounts} and methods in \citet{D16}). This results in an intensity of $\sim 12.2$~nW m$^{-2}$~sr$^{-1}$, which is 15\% larger but still compatible at the 1-sigma level with the output from our model of $\sim 10.6 \pm 1.7$~nW m$^{-2}$~sr$^{-1}$.

\subsubsection{Cosmic Optical (COB) and Infrared (CIB) Backgrounds}\label{sub:result_COBCIB}
The total  EBL intensity ($I_{bol}$, bolometric) integrated over all wavelengths and at any redshift, can be computed by doing: 

\begin{equation}\label{eq:Ibol}
 I_{bol} = \int_{\lambda_{min}}^{\lambda_{max}} \lambda I_{\lambda} (\lambda, z) \dfrac{d\lambda}{\lambda}
\end{equation}

\noindent where, in our case, $\lambda_{min} = 0.1~\mu m$ and $\lambda_{max} = 1000~\mu m$. Our calculations give an integrated local EBL intensity (Eq.\ \ref{eq:Ibol}) of 55.1$\pm$8.6~nW m$^{-2}$ sr$^{-1}$. This is roughly a factor of $\times$15 lower than the CMB intensity of $\sim$960~nW m$^{-2}$ sr$^{-1}$ \citep{Cooray16, Planck18}. We can divide the EBL into the cosmic optical background (COB, 0.1--8~$\mu$m) and cosmic infared background (CIB, 8--1000~$\mu$m). The integrated intensities of these two regions are shown in \autoref{fig:ebl_cob_cib} in comparison with the results from other works. Note that the COB is well constrained by all models within a factor of less than $\times$1.5; we get an integrated intensity of 23.1$\pm$4.0~nW m$^{-2}$ sr$^{-1}$. However, the data dispersion is larger in the CIB region, where we estimate the CIB intensity is 32.0$\pm$7.6~nW m$^{-2}$ sr$^{-1}$. We also reproduce this analysis by restricting our sample to only the MIPS+Herschel detected sources (\ie those galaxies for which the L(TIR) was directly measured). At lower redshifts, MIPS+Herschel sources are biased towards galaxies with a median stellar mass of $M_* \sim 10^{10}~M_{\odot}$, against $M_*\sim 10^{8}~M_{\odot}$ of the whole galaxy population at $z=0$. As a result, we conclude that the MIPS+Herschel resolved galaxies in our EBL sample contribute with the $\sim 40\%$ and $\sim 70\%$ of the integrated COB and CIB, respectively. This is in agreement with previous searches as those by \citet{Dole06}, where they also estimated a $\sim 70\%$ of resolved CIB by \emph{Spitzer}. These values are also shown in \autoref{fig:ebl_cob_cib} (COB value for MIPS+Herschel is below the intensity range of the plot).

\autoref{fig:ebl_mass_qsf} reveals the local COB and CIB contributions from quiescent versus SF galaxies according to UVJ diagram. SF galaxies dominates both the COB and CIB, making up $\sim 80\%$ and $\sim 90\%$ of the integrated EBL intensity, respectively. This SF relative contribution increases rapidly with redshift, while the quiescent contribution is mainly driven by passive galaxies at $z<1$, and becomes negligible at higher redshifts, where the quiescent population progressively disappears.

\subsubsection{Extragalactic background light redshift evolution}\label{sub:results_EBLz}
The evolution of the EBL with redshift is plotted in \autoref{fig:ebl_z} and compared with results from different models from the literature \citep{F10,D11,H12, Gilmore12} as well as from methodologies based on $\gamma$-ray attenuation measurements \citep{Fermi18}.

\autoref{fig:ebl_z} shows that most current EBL models are compatible in the UV/optical peak at $z<1$, however at higher redshifts different models diverge by about an order of magnitude in the UV. In the IR, there is a large disagreement between models at all redshifts, up to a factor of $\times$5 at the highest redshifts. These discrepancies led to significantly different optical depths at high redshifts for the lower energies ($\sim 50-100$~GeV) and at any redshift for the highest energies ($E\leq 10$~TeV, see Dom\'inguez et al., in preparation).

Furthermore, we can see in \autoref{fig:ebl_z} that our model generally agrees within $1\sigma$ with the latest results from $\gamma$-ray attenuation by {\em Fermi}-LAT \citep{Fermi18}. These $\gamma$-ray results tend to be on the higher end of our spectral intensities for the higher redshifts, although compatible within $2\sigma$. This situation was already noted by \citet{Fermi18} in comparison with other previous EBL models. Note that models tend to go on the lower bound of the $\gamma$-ray attenuation uncertainty band. Furthermore, interestingly, the $\gamma$-ray attenuation technique is sensitive to {\em all} light, even light that escapes galaxy surveys. Therefore, this discrepancy between galaxy surveys and $\gamma$-ray attenuation data, although not highly significant, could point to galaxy surveys' incompleteness in the NIR.

\begin{figure}
    \includegraphics[width=0.95\columnwidth]{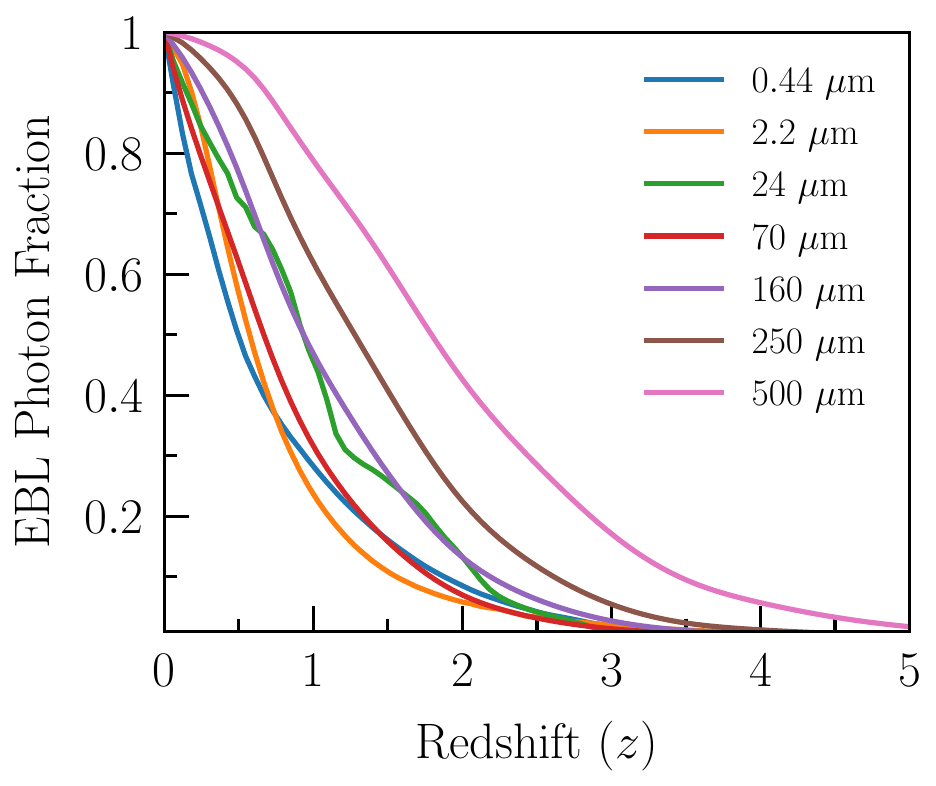}
\caption{The local EBL build up over redshift at different local wavelengths. The $y$-axis shows the fraction of light at that wavelength that was already in place at a given redshift, \ie produced at higher redshifts. For instance, more than 80\% of the local light at $500~\mu$m is produced at $z>1$, whereas only about 30\% of the light at $2.2~\mu$m come from these higher redshifts.}
\label{fig:buildup}
\end{figure}

We finally see in \autoref{fig:buildup} how photons populating the EBL at various wavelengths today were produced as a function of redshift. Photons in today's far-IR peak were emitted considerably earlier than photons in the optical peak. Our estimates show that approximately 80\% ($\pm$20\%) of the local $500~\mu$m light comes from $z>1$. This result, related with the existence of obscured starburst galaxies at high redshifts, is compatible with previous findings from the {\it Balloon-borne Large-Aperture Submillimeter Telescope} \citep{devlin09,marsden09}, {\it Spitzer} \citep{jauzac11}, and {\it Herschel} \citep{bethermin12}. However, only around 30\% ($\pm$5\%) of the local $2.2~\mu$m light comes from $z>1$, therefore the big majority of the local EBL at the optical peak was produced at $z<1$.

%% file: 5_summary.tex
\section{Summary}\label{sec:conclusions}
We present an empirical determination of the evolving EBL SED up to $z\sim 6$. In this study, the focus is on reducing current uncertainties on the EBL; that is, the FIR spectral region and the overall evolution with redshift. This is achieved by using unprecedented data from the CANDELS survey to consistently build the multiwavelength SED of approximately 150~000 galaxies from the UV to the FIR in five cosmological fields.

The UV-to-optical SED of every galaxy is reproduced following the work by \citet{Barro19}, whereas for the IR part we follow different approaches, according to whether the galaxy is detected in the FIR or not. We then compute the total galaxy luminosity density of the Universe over redshift by SED stacking and correcting for incompleteness, then we add up all this light to compute the evolving EBL. Our model generally reproduces (1) a diversity of FUV, NUV, optical, NIR and TIR luminosity density in the literature, (2) the CSFH, which is in agreement with the usual galaxy evolutionary hierarchical scenarios, (3) lower bound of the direct detection data, (4) galaxy counts and (5) latest results from $\gamma$-ray attenuation.

The improvements presented here will be beneficial for the derivation of robust optical depths for $\gamma$-ray photon fluxes and therefore, the interpretation of extragalactic VHE observations with {\it Fermi} LAT, current Cherenkov telescopes, and the upcoming Cherenkov Telescope Array (Dom\'inguez et al., in preparation).



